\documentclass[aps,prb,reprint,superscriptaddress]{revtex4-2}

\usepackage{graphicx}
\usepackage{dcolumn}
\usepackage{bm}
\usepackage{braket}
\usepackage[colorlinks]{hyperref}
\usepackage{color}
\usepackage[usenames,dvipsnames]{xcolor}
\usepackage{comment}
\usepackage[export]{adjustbox}
\usepackage{float}

\emergencystretch=\maxdimen
\begin{document}


\title{Thermal self-oscillations in monolayer graphene coupled to a superconducting microwave cavity}

\author{Mohammad Tasnimul Haque}
\affiliation{Low Temperature Laboratory, Department of Applied Physics, Aalto University, P.O. Box 15100, FI-00076 Espoo, Finland}
\author{Marco Will}
\affiliation{Low Temperature Laboratory, Department of Applied Physics, Aalto University, P.O. Box 15100, FI-00076 Espoo, Finland}
\author{Alexander Zyuzin}
\affiliation{QTF Centre of Excellence, Department of Applied Physics, Aalto University, P.O. Box 15100, FI-00076 Aalto, Finland}
\author{Dmitry Golubev}
\affiliation{QTF Centre of Excellence, Department of Applied Physics, Aalto University, P.O. Box 15100, FI-00076 Aalto, Finland}
\author{Pertti Hakonen}
\affiliation{Low Temperature Laboratory, Department of Applied Physics, Aalto University, P.O. Box 15100, FI-00076 Espoo, Finland}
\affiliation{QTF Centre of Excellence, Department of Applied Physics, Aalto University, P.O. Box 15100, FI-00076 Aalto, Finland}

\vspace{10pt}

\begin{abstract}
  Nonlinear phenomena in superconducting resonator circuits are of great significance in the field of quantum technology. We observe thermal self-oscillations in a monolayer graphene flake coupled to Molybdenum-Rhenium superconducting resonator. The graphene flake forms a SINIS junction coupled to the resonator with strong temperature dependent resistance. In certain conditions of pump power and frequency, this nonlinearity leads to thermal self-oscillations appearing as sidebands in cavity transmission measurements with strong temperature dependence and gate tunability.
  The experimental observations fit well with theoretical model based on thermal instability. The modelling of the oscillation sidebands provides a method to evaluate electron phonon coupling in disordered graphene sample at low energies. 

\end{abstract}

\maketitle

\section{Introduction}
Many phenomena in nature are self-oscillatory \cite{Jenkins2013}.
In electrical circuits, relaxation oscillators make up a basic class of such self-oscillating systems. They relate an intrinsic RC time scale to a frequency which is easy to determine and to convert to actual circuit parameters. In a properly designed thermal system, thermal time constants and the underlying parameters can be determined in a similar fashion. In this work, we show that a microwave cavity connected to a 2D graphene object forming a superconductor-insulator-normal conductor-insulator-superconductor resistor ($R_{{\rm SINIS}}$) can be employed to determine thermal time constants related to the electron-phonon coupling using thermally induced self-oscillation in the cavity photon occupation number. The energy of the cavity is dissipated to the graphene resistance, and the rate of this dissipation is strongly dependent on the temperature of the graphene membrane. The temperature of graphene is governed by the electron-phonon coupling that carries away the dissipated Joule heating arising from the decay rate of the cavity photons. Owing to the exponential non-linearity of the $R_{{\rm SINIS}}$ resistance, self-oscillations in the range of 0.5 -100 MHz appear. At low temperatures below 0.5 K, the oscillation frequency can be employed to determine the electron-phonon coupling having a $T^3$ temperature dependence of the heat flow. 

Such a power law can be explained invoking the electron-phonon disorder-assisted scattering processes.
The $T^3$ dependence agrees with experiments performed at temperatures higher than the Bloch-Gr\"uneisen temperature \cite{Betz2012,Laitinen2014} where the findings were assigned to disorder assisted electron-phonon scattering events (supercollisions) \cite{Song2012}.
Our investigation was performed for disordered graphene at low temperatures well below the Bloch-Gr\"uneisen temperature, where $T^3$ dependence might be expected as well \cite{Clerk2012}. 

\subsection{Electron-phonon coupling}
The low electron density in graphene near the Dirac point 
 results in a relatively weak electron-acoustic phonon coupling, smaller than in a conventional
metal \cite{Kubakaddi2009,Tse2009,Bistritzer2009,Viljas2010,Betz2012,Suzuura2002,Katsnelson2012}. 
The reason for that is the restriction of energy transfer in the momentum conserved electron-phonon collisions for systems with small Fermi surface \cite{Katsnelson2012}.
Indeed, the maximum amount of momentum transfer for carriers at the Fermi level is twice the Fermi momentum $2 k_{\mathrm{F}}$, which facilitates energy transfer by $ 2v_{\mathrm{s}} \hbar k_{\mathrm{F}}$ per one phonon, where $v_{\mathrm{s}}$ is the speed of sound and $\hbar$ is the Planck constant. This energy defines the Bloch-Gruneisen temperature $T_{\mathrm{BG}} = 2v_{\mathrm{s}} \hbar k_{\mathrm{F}}/ k_{\mathrm{B}} $, where $k_{\mathrm{B}}$ is the Boltzmann constant, above which only a fraction of phonons may scatter from electrons residing in the thermally activated energy window. This has been observed, for example, in resistance vs. temperature measurements in graphene \cite{Chen2008,Efetov2010}.
As the temperature drops below $T_{\mathrm{BG}}$ the linear-$T$ dependence of the heat flow switches to $T^4$.

However, the momentum conservation constrains can be relaxed for processes involving electron scattering from two flexural phonons or defect assisted electron-phonon scattering \cite{Song2012,Graham2012,Betz2012a,Graham2013,Clerk2012,Mariani2010,Castro2010,Katsnelson2012}.
In the latter, the interaction processes are dressed by the electron scattering from impurities or dynamic ripples. 
Here one can introduce a characteristic temperature associated with the disorder scattering \cite{Song2012, Clerk2012}, which is defined as
$
T_{\mathrm{dis}}= 2\pi \hbar v_s/k_{\mathrm{B}} L_{\mathrm{e}}< T_{\mathrm{BG}}$, where $L_{\mathrm{e}}$ denotes the mean free path of the charge carriers. Assuming weakly screened electron-phonon interaction, with the increase of temperature from $T \ll T_{\mathrm{dis}}$ via
$T_{\mathrm{dis}} <T < T_{\mathrm{BG}}$ to $T_{\mathrm{BG}}<T$, the energy loss power changes from $T^3$ to $T^4$ and then back to $T^3$, respectively \cite{Song2012, Clerk2012}.
Experiments have indicated that such electron-phonon-impurity interference events dominate over normal electron-phonon scattering in regular graphene samples, both  non-suspended and suspended \cite{Betz2012, Laitinen2014}. 

Generally, the heat flow due to electron-phonon coupling $P_{\mathrm{e-ph}}$ from charge carriers to the lattice can be expressed as a power law 
\begin{equation}
  P_{\mathrm{e-ph}}=\Sigma A (T^{\gamma} -T_{\mathrm{ph}}^{\gamma}) ,
\end{equation}
where $\Sigma$ is the coupling constant, $A$ is the area of the graphene flake,
$T$ denotes the electron temperature, $T_{\mathrm{ph}}$ specifies the phonon temperature, and $\gamma $ is the characteristic exponent \cite{Giazotto2006}.

 In our experiments, the charge density in the suspended part has been varied over $|n| < 1.8 \cdot 10^{11}$ cm$^{-2}$, while the residual charge density $n_0 \simeq 1.5 \times 10^{10}$ cm$^{-2}$. Thus, $T_{\mathrm{BG}} = 3.8 - 23$ K for longitudinal acoustic phonon dispersion relation with sound speed $v_s=2 \times 10^4 $ m/s. However, the electron-phonon heat flow remains small in the suspended part because of its minor area and small $|n|$.

The contact regions having a charge density $n_c \sim 3\times 10^{12}$ cm$^{-2}$ fully dominate thermal flow from electrons to phonons in our samples. This value of $n_c$ corresponds to $T_{\mathrm{BG}} \simeq 90$\,K which well exceeds the estimate $T_{\mathrm{dis}} \simeq 30$\,K obtained from a typical mean free path $\sim 30$\,nm for graphene on rough surfaces. Thus, our experiments  probe the graphene electron-phonon coupling in the disordered limit, 
and our data at an average temperature $\braket{T} < 10 $ K deal with the regime of $T \ll T_{\mathrm{dis}} < T_{\mathrm{BG}}$, where the electron-acoustic phonon scattering in the weak electronic screening leads \cite{Clerk2012} to $\gamma = 3$ with $\Sigma = \frac{2\zeta(3) D^2|E_{\mathrm{F}}|k_{\mathrm{B}}^3}{
\pi^2\rho_{\mathrm{M}}\hbar^4v_\mathrm{F}^3
v_\mathrm{s}^2 L_{\mathrm{e}}}$, where $D$ is the deformation potential of graphene, $\rho_{\mathrm{M}}$ is the mass density, $E_{\mathrm{F}}$ is the Fermi energy. Using the strength of the deformation potential as a fitting parameter, a broad range of $D\simeq10 \dots 70$ eV has been obtained from the experiments probing the electron-phonon coupling \cite{Betz2012,Betz2012a,Fong2012,Fong2013,Han2013,Vora2014,McKitterick2016,ElFatimy2019}.

\subsection{Principle of thermal self-oscillation}
Thermal hysteresis has been found in superconducting SNS junctions \cite{Courtois2008}. The retrapping current is lowered because the temperature $T$ dependent supercurrent is smaller in the state with Joule heating due to the normal current. Thermal hysteresis appears also in SINIS structures without any supercurrent. This arises from the strongly non-linear resistance of the device. Owing to the exponential reduction of quasiparticles with lowering temperature, the resistance becomes proportional to $\exp(\Delta/k_B T)$ where $\Delta$ denotes the energy gap of the superconducting electrodes. When the device resistance $R_{\mathrm{ SINIS}}$ is voltage biased, the temperature will be balanced to a value at which the Joule heating $V^2/ R_{\mathrm{ SINIS}}$ will be compensated by the electron-phonon coupling heat flow $P_{\mathrm{e-ph}}(T)$, provided that the quasiparticle heat transport in the superconductor can be neglected. 

\begin{center}
\begin{figure}[htb]
\includegraphics[width=0.9\linewidth]{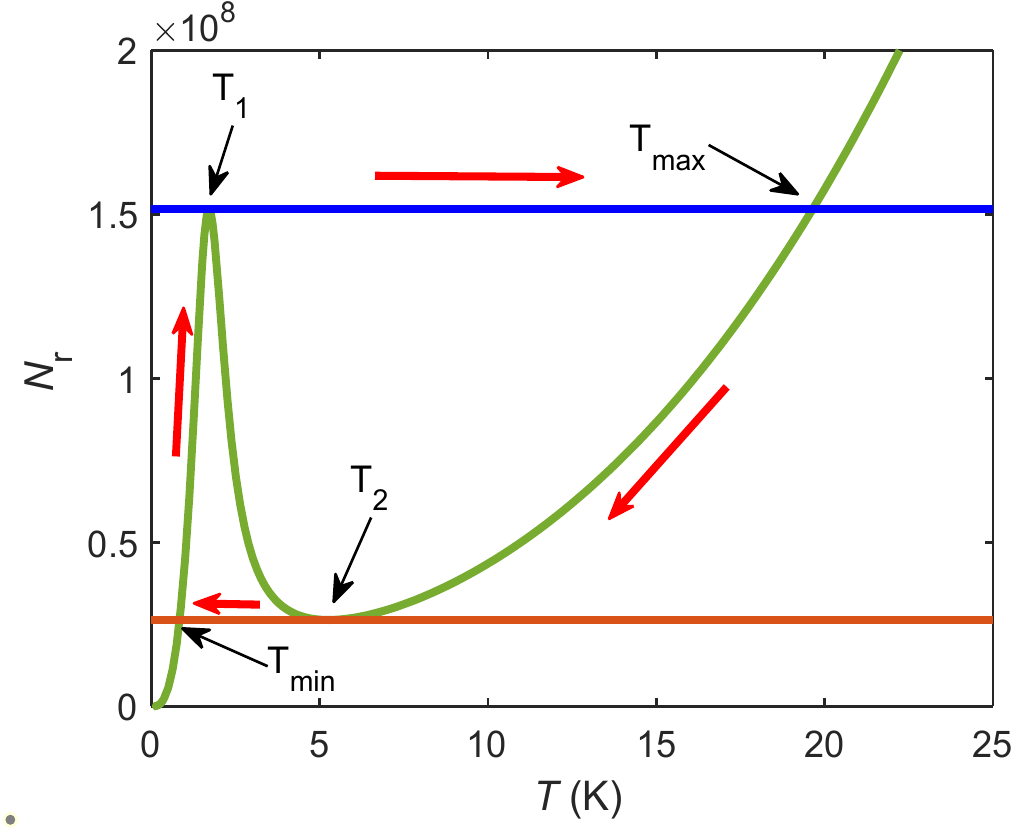}
\caption{Illustration of the thermal hysteresis loop in a microwave cavity with dissipation governed by a SINIS structure; the employed thermal model is described in App. \ref{a1}. The y-axis denotes number of microwave photons $N_r(T)$, given by Eq. (\ref{NrT}), which provides the voltage for Joule heating. The simulation parameters of the system: the resistance of SINIS structure is given by Eq. (\ref{Rqp}) with $R_0=3$ M$\Omega$,  $R_N=500$ $\Omega$, $\Delta=1.76 k_B T_C$, and $T_C=9$ K, while the electron-phonon coupling is defined by $\gamma=3$ and $A\Sigma=9.2\times 10^{-11}$ W/K$^\gamma$. The cavity has a frequency $\omega_0/2\pi = 5.382$ GHz, characteristic impedance is set to $Z_C=100$ $\Omega$ and the simulation is performed at $T_0=20$ mK.}
\label{Fig:Nr_T}
\end{figure}
\end{center}

Owing to exponential temperature dependence of $R_{\mathrm{ SINIS}}$, the Joule heating may increase faster than what can be compensated by $P_{\mathrm{e-ph}}(T)$ and a thermal run away takes place. Once $T \simeq \Delta/k_B$,  however, the increase in heating becomes limited by saturation of sample conductance $1/ R_{\mathrm{ SINIS}}$, and a new stable temperature can be obtained. Similarly, when lowering voltage, there will be an unstable range of temperature, and a thermal hysteresis loop is formed. In our case, the heating voltage is governed by the number of photons in a microwave cavity to which $R_{\mathrm{ SINIS}}$ is connected. The jump in $T$ causes a jump in quality factor that governs the stationary number of quanta in the cavity. This leads to bistability of the cavity, and self-oscillations in a range of drive powers of the microwave cavity. A hysteresis loop as a function of the number of microwave quanta (voltage$^2$) is illustrated in Fig. \ref{Fig:Nr_T}. The speed at which the hysteresis loop is traversed depends on thermal characteristics of the SINIS structure, so that the self-oscillation frequency can be employed  for determination of the electron-phonon coupling in graphene at low energy.

The structure of the hysteresis loop arises from two coupled differential equations, one governing the decay rate for cavity photons $dN_r(T)/dt$ and one dealing with the heat balance for the rate of change of electron temperature $dT/dt$ in the graphene SINIS structure. Owing to large ratio of $\Delta/k_B T$, we may neglect the electronic heat transfer via the superconducting leads and can write: 
\begin{eqnarray}
\frac{dN_r}{dt} &=& -(\kappa_{\rm in}+\kappa_{\rm out}+\kappa_G(T))(N_r-N_{st}(T)),
\nonumber\\
C_G(T)\frac{dT}{dt} &=& - A\Sigma(T^\gamma - T_0^\gamma) + \hbar\omega_r \kappa_G(T) N_r.
\label{rate_syst}
\end{eqnarray}
Here $\kappa_{\rm in}$, $\kappa_{\rm out}$, and $\kappa_G(T)$ denote the cavity decay rates due to input, output, and graphene dissipation, respectively, $T_0$ is the bath temperature, $N_r(T)$ specifies the photon occupation number of the microwave cavity and $N_{\rm st}$ is the steady state number of photons; $C_G$ denotes the heat capacity of graphene which is assumed to be negligibly small near Dirac point at low temperature. By setting $C_G=0$, these equations can be solved straightforwardly as shown in App. \ref{a1}. Crudely, the self-oscillation takes place between two cavity states, a high-$Q$ state ($Q \sim 5000$) and a low-$Q$ state ($Q \sim 20$), which leads to a slow build up of photon occupation in the cavity at low $T$, followed by a quite fast release of quanta during the low-$Q$ operation at high $T$ (see Figs. \ref{Nr(t)} and \ref{T(t)} in Sect. \ref{results}).

In Ref. \citenum{Segev_2007}, a closely related emergence of thermal self-oscillations in superconducting resonators have been reported. In their device, a NbN stripline ring resonator is integrated with a superconducting microbridge to enhance nonlinear effects. When pumped with a monochromatic signal, the microbridge acts as a hotspot, oscillates between superconducting phase and normal conducting phase resulting in thermal self-oscillations. Self-oscillations arising from thermally-induced instabilities have also been reported in other nanodevices such as in thermo-optic nanocavities \cite{madiot_vibrational_2021} and carbon nanotube based NEMS resonators \cite{urgell_cooling_2020}, as well as in optical parametric oscillators \cite{Suret_2000}.

\section{Experiment}
\begin{center}
\begin{figure}[htb]
\includegraphics[width=0.99\linewidth]{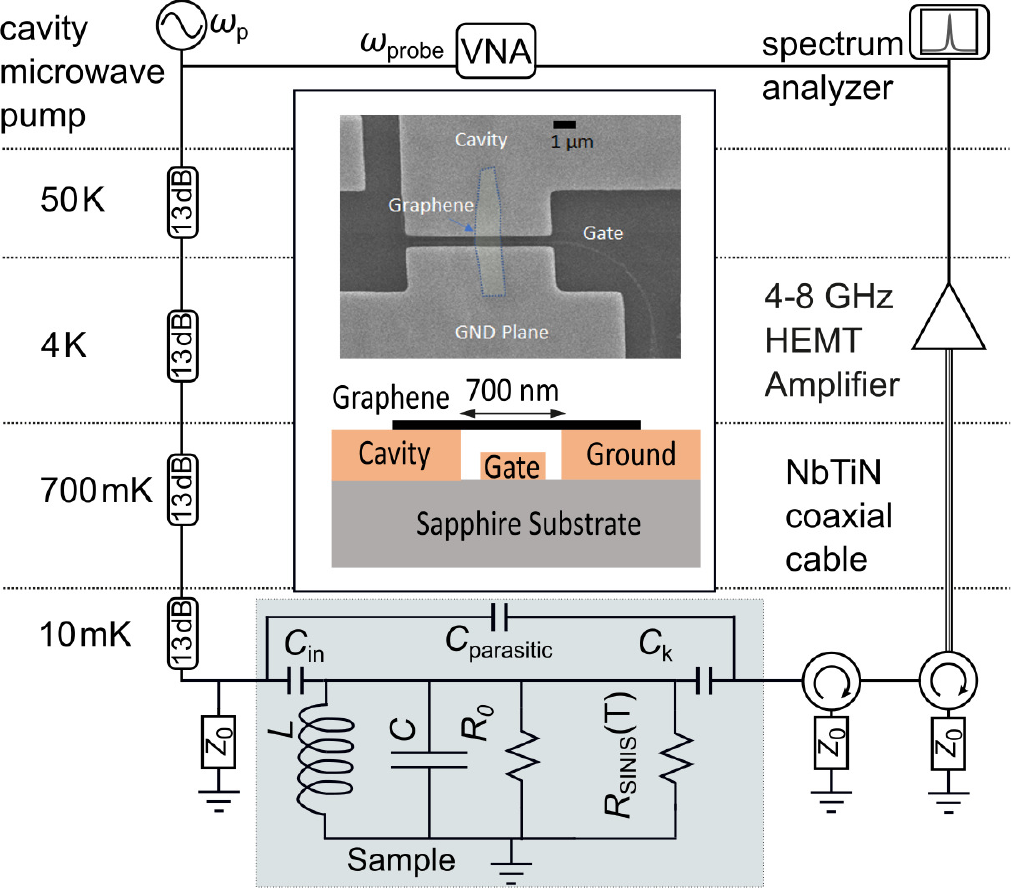}
\caption{Microwave measurement scheme and an equivalent model of the $\lambda/2$ cavity employed in the experiments. The inset displays a scanning electron microscope (SEM) image of the graphene sample extending over a $700$\,nm gap between the center conductor of the cavity and the ground plane; the graphene element is suspended over a local gate as indicated by the cross sectional view below the SEM image. For emission measurements the cavity is driven by a pump signal on the cavity resonance frequency dependent on the pumping power. In transmission measurements, additionally, a probe signal is set to the side peak frequency governed by the frequency of thermal oscillations in the sample. For details, see text.}
\label{setup}
\end{figure}
\end{center}

Our sample and the employed experimental setup is depicted in Fig. \ref{setup}. The sample consists of a monolayer graphene flake coupled to a superconducting microwave cavity (see the inset in Fig. \ref{setup}). An exfoliated monolayer graphene flake of area $A=40$ $\mu$m$^2$ (total length $L_T \simeq 16\,\mu$m, width $W \simeq 2.5\,\mu$m) is deposited onto the chip using dry transfer method in such a way that the flake is suspended over the local gate while making contacts with the center strip of the cavity transmission line and the ground plane.
The gap between the cavity transmission line and the ground plane is 700 nm, which defines the length of the suspended graphene part.

The superconducting cavity was made of Molybdenum-Rhenium (60:40) alloy superconductor on a r-cut sapphire substrate. Molybdenum-Rhenium (MoRe) was chosen because it has a high critical temperature $T_C \sim 9$\,K, a high critical magnetic field $B_C \sim 8$\,T and it makes transparent contacts with graphene. MoRe film of 300 nm thickness was first co-sputtered on a sapphire substrate at 750C and then the cavity with local gates were patterned with two step electron beam lithography. The cavity was designed to have a characteristic impedance of $Z_C \sim 100$\, $\Omega$ with an input coupling capacitor $C_{in} = 0.5$\,fF and an output coupling capacitor near graphene $C_K = 2.21$\,fF. The cavity is formed as a $\lambda/2$ transmission line resonator with two voltage anti-nodes situated at the two ends of the line and a voltage node at the center of the cavity length. For DC voltage biasing, a superconducting broadband reflective T filter \cite{HaoYu2014} is connected to the cavity at the voltage node. This allows to apply DC voltages into the center trace of the cavity without loss of quality factor. Two 50\,$\Omega$ RF transmission lines are connected for microwave measurements. On the input side, there is 52 dB attenuation on the RF line and additionally, a parasitic capacitive shunt path as shown in Fig. \ref{setup}. On the output side, the signal is transmitted through two circulators and then amplified by a 4-8 GHz HEMT amplifier mounted at the 4K stage. The two circulators protect the sample from noise coming back from the HEMT amplifier. The cavity can be modeled as a parallel RLC circuit with an additional resistor $R_{\mathrm{SINIS}}$, see the equivalent circuit in Fig. \ref{setup}. The $Q$ factor for the equivalent resonant circuit is given by $Q = \omega_0 R C$ where $1/R = 1/R_{\mathrm{SINIS}}+1/R_0$ denotes the total conductance of graphene (G) and the cavity subgap resistance, respectively, while $C \approx 0.47$\,pF is calculated from the geometry of the cavity. As described in detail in App. \ref{SuppExp}, the recorded transmission spectra are fitted with Eq. (\ref{Cavity_fiTquation}) to estimate temperature dependence of the $Q$ factor of the cavity. The fits indicate that both $R_{\mathrm{SINIS}}$ and $R_0$ depend on $T$, but the latter can be typically taken as constant.

The sample was mounted on the mixing chamber of a Bluefors LD400 dry dilution refrigerator with $T$ = $10$\,mK base temperature. First we characterised the DC response of the sample as a function of gate voltage $V_g$. The charge density $n=C_g V_g$ was obtained using parallel plate approximation for the capacitance $C_g=0.08$\,fF. From the conductance measurement $G(n)$, the Dirac point having residual charge carrier density $n_0=10^{10}$ cm$^{-2}$ is located at gate voltage $V_g = 1.7$\,V while majority of the measurements were carried out at gate voltage $V_g = 0$\,V corresponding to a charge density of $n = -4.7 \times 10^{10}$\,cm$^{-2}$ (chemical potential $\mu_0=25$ meV). At this gate  bias point, we estimated the apparent mobility of the graphene flake as $\mu =\frac{W}{L} (G-G_{\mathrm{min}})/n e \simeq  35000$\,cm$^{2}$/Vs where $L=0.7$\,$\mu$m is the length of the suspended part between the electrodes. 
However, this mobility is influenced by the enhanced contact resistance due to superconductivity of the contacts.
Accounting for the contacts, this mobility corresponds to a mean free path $L_e=100$\,nm which agrees with characteristics of similar suspended samples \cite{Kumar2018,Kamada2021}. In App. \ref{SuppExp} we display our data on $G(n)$ measured at $T=4$\,K. The mean free path in graphene on top of MoRe conductor is estimated to be $L_e=20-30$\,nm due to roughness-induced strain variation \cite{Cuoto2014}.

\begin{center}
\begin{figure}[htb]
\centering
\includegraphics[width=0.975\linewidth]{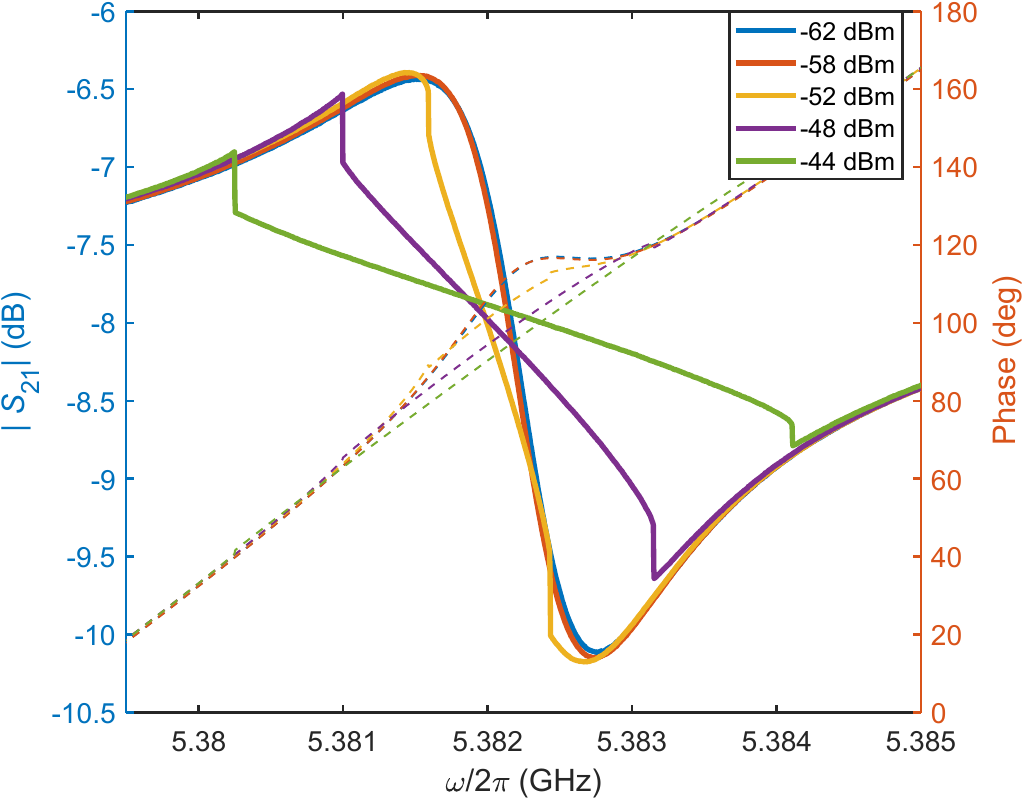}
\caption{Magnitude (solid lines) and phase (dotted lines) of transmission $S_{21}$ through the cavity for five input powers $P$ indicated in  the inset. The cavity transmission signal has Fano resonance shape, but at probe powers $P > P_{\mathrm{th}}$, thermal oscillations appear and the shape of the resonance starts to deviate from normal resonance shape with appearance of abrupt steps in magnitude of $S_{21}$ and a strong reduction of the effective $Q$ factor in the resonance region of the signal.}
\label{cavity_vs_power}
\end{figure}
\end{center}

Fig. \ref{cavity_vs_power} displays the measured  transmission signal $S_{21}$ through the cavity at a few microwave powers. Owing to interference with a shunt signal due to parasitic capacitance,  the recorded transmission spectrum has Fano resonance shape. The line width of the cavity resonance at small signal powers corresponds to a quality factor of $ Q= 5000$ at the base temperature. The shape of the resonance was employed to deduce the temperature-dependent value of $R_{\mathrm{SINIS}}(T)$ for the graphene resistor at the cavity frequency $\omega_0/2\pi$.

Above a certain threshold probe power, the shape of the resonance starts to break away from usual shape and small jumps in transmission appear both above and below the resonance frequency. Between the steps, the slope of the transmission signal becomes smaller, which indicates a reduction in the $Q$ factor upon the step. Such a stepwise change in the $Q$ factor can be considered as a sign of thermal runaway. With increasing probe power, the separation of the steps becomes larger, which indicates thermal runaway further away of the resonance. This suggests that there is a critical voltage, because this threshold voltage, or equivalently the critical number of quanta $N_c$, can be reached further away from the resonance at large drives. Crossing of the critical value $N_c$ is assigned to thermal runaway and the onset of thermal self-oscillation.

\begin{center}
\begin{figure}[htb]
\centering
\includegraphics[width=\linewidth]{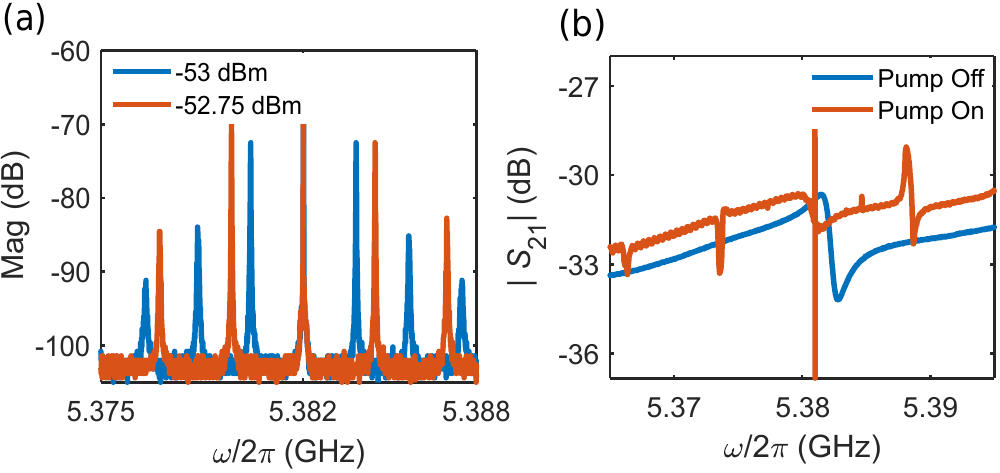}
\caption{(a) Emission spectrum emerging from the sample under thermal self-oscillations. The cavity is pumped at resonance frequency above the threshold power $P_{\mathrm{th}}=-53.2$\,dBm: $P = -53$\,dBm (blue) and $P= -52.75$\,dBm (red). The sideband peaks shift to higher frequencies with increasing pump power. (b) Comparison of transmission spectra taken with the pump drive off (blue) and with the pump on at the cavity resonance frequency (red). The pumping frequency is selected dependent on the  power so that the pump remains around the resonance center. }
\label{EmissionSpectrum}
\end{figure}
\end{center}

We have made two kinds of studies on the thermal oscillations. In emission spectrum measurements, we pump the cavity with a monochromatic RF signal at cavity resonance frequency $\omega_0/2\pi = 5.382$\, GHz and record the transmitted signal with a spectrum analyzer. If pumped above threshold power $P_{\mathrm{th}} = -53.2$\,dBm, thermal self-oscillation sidebands appear in the emission spectra whose peaks shift to higher frequency separation with increasing power, see Fig. \ref{EmissionSpectrum}(a). In transmission measurements, in addition to pump signal, we apply a low power probe signal ($P_{\mathrm{probe}} = -110$\,dBm) from a VNA and record $S_{21}$ parameter. Fig. \ref{EmissionSpectrum}(b) shows comparison of cavity transmission signals without any pumping and with pumping above $P_{\mathrm{th}}$. When pumping is on, we observe sidebands with Stokes and anti-Stokes like features. Similar sidebands have been observed in optomechanical systems and utilized for amplification and cooling. 

An additional point to observe concerning Fig. \ref{EmissionSpectrum}(b): when the pump is on, due to its substantial power, the cavity heats up and its resonance frequency shifts owing to a change in kinetic inductance $L_k$ of the MoRe superconductor. Consequently, the pump frequency needs to be adjusted so that it stays at the resonance. Fig. \ref{cavity_vs_temp} in App. \ref{SuppExp} displays the cavity resonance shift due to $L_k$ as a function of $T_0$. The observed shift in resonance frequency with pump on in Fig. \ref{EmissionSpectrum}(b) indicates that the effective electronic temperature in the cavity is $T\simeq 1.7$\,K .

\section{Results \& Discussions} \label{results}

\begin{center}
\begin{figure}[tb]
\centering
\includegraphics[width=0.9\linewidth]{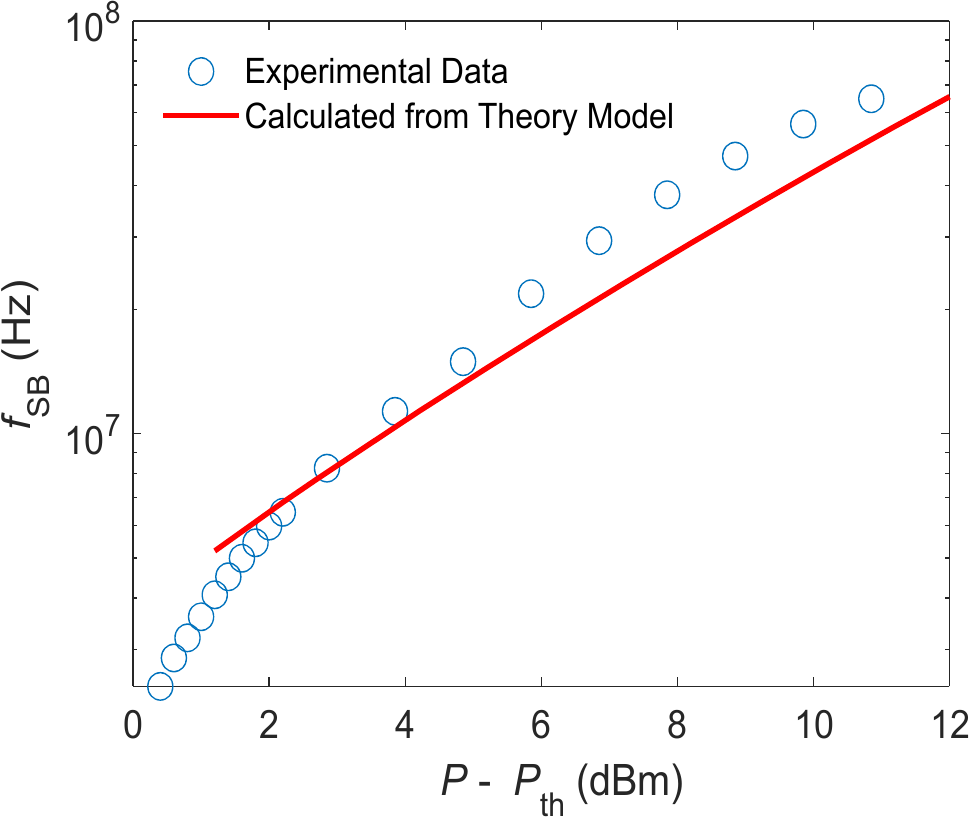}
\caption{Shift of the sideband frequency $f_{\mathrm{SB}}$ as a function of pump power difference from the critical pump, $P_{\mathrm{th}} = -53.2$ dBm. 
The solid red curve displays the simulated behavior obtained from the coupled equations in Eq. (\ref{rate_syst}).}
\label{Sidebands_vs_P}
\end{figure}
\end{center}

Our experiments reveal thermal oscillations over a certain range of pumping powers, under which the drive keeps the system in its bistable regime. First, one has to reach the temperature of thermal runaway, which is achieved above a threshold power $P_{\mathrm{th}}$, corresponding to the threshold number of quanta $N_c$ in the cavity. Second, the power has to be smaller than a maximum value $P_m$ in order to facilitate reaching the minimum of $N_r(T)$ curve during the decay of photon occupation in the low-$Q$ state. This range of powers, $P_{\mathrm{th}}=-53.2$\,dBm $ < P < P_m = -42$\,dBm in the experiment, is seen in Fig. \ref{Sidebands_vs_P} which illustrates the power dependence of the thermal sideband frequency $f_{\mathrm{SB}}$ at the cryostat base temperature 10\,mK. The $f_{\mathrm{SB}}$ data in Fig. \ref{Sidebands_vs_P} indicate a strong increase of the self-oscillation frequency from about 2 MHz to 70 MHz. 

The frequency of self-oscillations in Fig. \ref{Sidebands_vs_P} is given in terms of the excess power above the threshold power $P_{\mathrm{th}}$. The behavior is nearly linear. 
Assuming that the oscillation frequency is governed by the ramp up in the number of quanta with $\kappa=\kappa_{\rm in} + \kappa_{\rm out}+\kappa_G(T)={\rm const}.$, and that the stationary number of quanta $N_{st} \gg N_c$, the self oscillation frequency may be approximated by $f_{\mathrm{SB}} \sim \kappa N_{st}/N_c$. As long as we may neglect any increase in $\kappa$ and the ensuing change in $N_{st}$, the frequency $f_{\mathrm{SB}}$ would be proportional to $P_{\rm in}$, which is in qualitative agreement with the observed behavior. This approximation, however, is only valid at intermediate powers and full solution of Eqs. (\ref{rate_syst}) is needed for proper analysis.

The solid red curve in Fig. \ref{Sidebands_vs_P} displays the simulated behavior obtained from the coupled equations (\ref{rate_syst}). The overall power dependence from our model, using constant subgap resistance $R_0$ in the cavity, coincides well with the experimental data. The following parameters are used for the simulation: the resistance of SINIS structure is given by Eq. (\ref{Rqp}) with $R_0=3$ M$\Omega$,  $R_N=500$ $\Omega$, $\Delta=1.76 k_B T_C$, and $T_C=9$ K while the electron-phonon coupling is defined by $\gamma=3$ and $\Sigma A=9.2\times 10^{-11}$ W/K$^\gamma$. This value of $\Sigma A$ would become reduced by 2\% if hot quasiparticles entering the cavity would be taken into account (see below). 

\begin{center}
\begin{figure}[htb]
\centering
\includegraphics[width=0.9\linewidth]{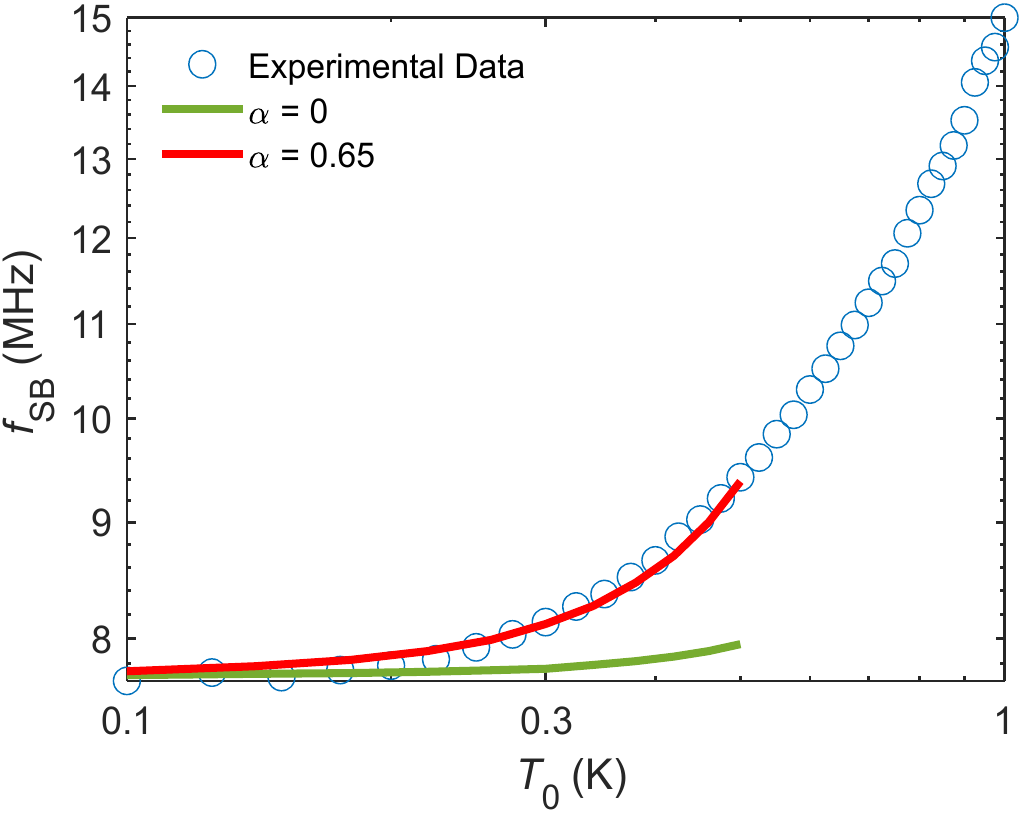}
\caption{Shift of thermal oscillation sideband frequency as a function of temperature. Blue circles are experimental data and solid lines are calculated from theory model with an addition of temperature dependence of the subgap resistance in the form $R_0 = R_{00}$exp$(\Delta/k_B \sqrt{T_0^2 + (\alpha T)^2})$ with $\alpha= 0$ for green and $\alpha= 0.65$ for red line. For this fit, we employ electron-phonon coupling constant $\Sigma A = 9\times10^{-11}$\,W/K$^3$ obtained form the data in Fig. \ref{Sidebands_vs_P}. 
}
\label{Sidebands_vs_T}
\end{figure}
\end{center}

Fig. \ref{Sidebands_vs_T} displays the self-oscillation frequency $f_{\mathrm{SB}}$ as a function of temperature. The measurement power $P = -50.5$ dBm is approximately in the middle of the logarithmic power scale $P_{\mathrm{th}} \dots P_m$. The increase in $f_{\mathrm{SB}}$ in Fig. \ref{Sidebands_vs_T} arises from a change in the electron-phonon coupling in the sample with increasing $T$ along with drive power. In general, the temperature dependence of $f_{\mathrm{SB}}$ reflects the $T$ dependence of $P_{\rm e-ph}$, either arising directly through the coupling or due to extra dissipation by hot quasiparticles injected to the superconducting cavity from heated graphene. At low $T < 0.5$\,K, the measured $f_{\mathrm{SB}}$ follows $T^3$ dependence of $P_{\rm e-ph}$ very closely.

The solid green curve in Fig. \ref{Sidebands_vs_T} displays the simulated behavior for $f_{\mathrm{SB}}(T_0)$ obtained from our model with $R_0=\mathrm{const}$. The weak  dependence on $T_0$ suggests that there is an additional factor  for the cavity photon number relaxation that becomes strengthened with $T_0$. Obviously, the larger $T_0$ is, the more quasiparticles there will be in the superconductor of the cavity resonator. The density of quasiparticles will also enhance with the electronic heating of graphene due to deposition of photon energy. We approximate the subgap resistance of the cavity with extra injected  quasiparticles by $R_0^* = R_{00} \mathrm{exp}(\Delta/k_B \sqrt{T_0^2 + (\alpha T)^2})$ where $R_{00}= 10 \Omega$ and $\alpha$ is a tunable parameter. This $R_0^*$ with $\alpha=0.65$ yields a good agreement with the measured data as seen from the red curve in Fig. \ref{Sidebands_vs_T}. The obtained good agreement in both dependencies in Figs.  \ref{Sidebands_vs_T} and \ref{Sidebands_vs_P} allows us to determine the effective electron-phonon coupling value $\Sigma A=9.2\times 10^{-11}$\,W/K$^3$ from the thermal oscillation with high confidence. 

Owing to fast electronic heat transport in graphene, the effective area for heat relaxation is the total area of the sample $A=40$ $\mu$m$^2$. Thus, we obtain $\Sigma=2.2$\,W/m$^2$K$^3$. This value is slightly larger than the result of Refs. \onlinecite{Borzenets2013,ElFatimy2019}, though it is hard to estimate the actual area for their complex sample geometries. Part of the difference with Ref. \onlinecite{Borzenets2013} may also arise due to the different contact structure: in our device we have a weak overlay contact of graphene to MoRe, instead of the commonly-used evaporated metal on graphene structure.    

Density functional theory (DFT) calculations on graphene/metal contacts have been performed in Ref. \onlinecite{Khomyakov2009} for several common metals. According to these calculations, doping of graphene in the contact region depends on the work function difference $\Delta E_F = W_M-W_G^*$, where $W_M$ and $W_G^*$ denote the work function of the metal and that of the graphene in contact with the metal, respectively \cite{Khomyakov2009,Laitinen2016}. In gold and silver contacts, for example, $|\Delta E_F| \sim 0.2-0.3$\,eV, whereas experiments on gold indicate even slightly larger doping $\Delta E_F= -0.35$\,eV \cite{Nagashio2009}. Using $E_F=0.3$\,eV, $D=50$\,eV, $L_e=25$\,nm, $v_F=10^6$\,m/s, and $v_s=2\times10^4$\,m/s, we obtain $\Sigma=2.1$\,W/m$^2$K$^3$, which is close to the measured result. Note that this agreement between experiment and theory is dependent on the amount of doping by the contact metal which is poorly known at present for MoRe contacts.

\begin{center}
\begin{figure}[tb]
\includegraphics[width=0.9\linewidth]{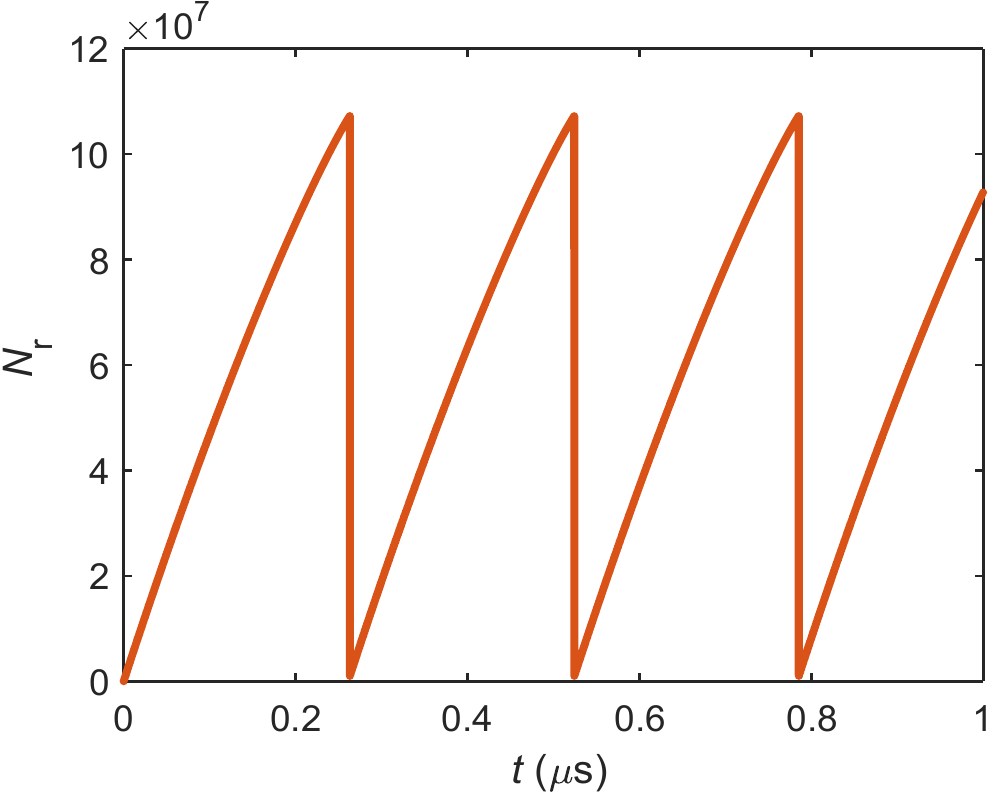}
\caption{Time dependence of the number of photons in the resonator, $N_r(t)$, obtained by numercally solving Eqs. (\ref{dNdt}).
The system parameters are the same as in Fig. \ref{Fig:Nr_T}. In addition, we have set $P=-53$ dBm for drive power and $\mu=10$ meV for
the chemical potential of graphene. }
\label{Nr(t)}
\end{figure}
\end{center}

Our thermal oscillation model yields a nearly saw-tooth pattern for the time dependence of the number of photons in the resonator $N_r(t)$. The obtained theoretical pattern $N_r(t)$ is illustrated in Fig. \ref{Nr(t)} starting from zero quanta in the cavity; a steady state oscillation is obtained right from the first cycle. The oscillation frequency is approximately 3.8 MHz, and it corresponds well to the measured frequency at $P=-53$\,dBm. The build up of occupation is gradual with a time constant on the order of 0.45 $\mu$s while the decrease is quite abrupt on the time scale of the figure. This saw-tooth pattern governs the emission of microwave quanta under the self-oscillation conditions. 

\begin{center}
\begin{figure}[htb]
\includegraphics[width=0.9\linewidth]{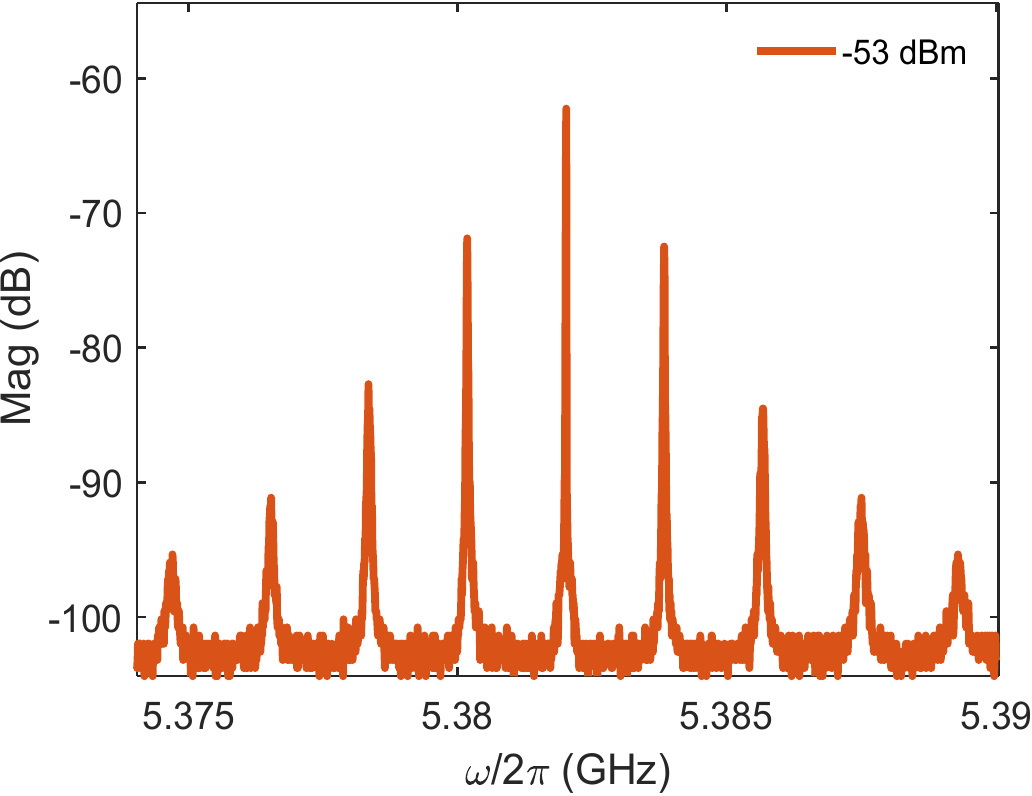}
\caption{Emission spectrum of thermal oscillations at pumping power $P = -53$ dBm applied at $f_p=5.382$ GHz; this power corresponds to the calculated trace in Fig. \ref{Nr(t)}. The areas of these thermal-oscillation-driven side band peaks compare well with Fourier power spectrum of a sawtooth pattern. See text for details.}
\label{EmissionStrong}
\end{figure}
\end{center}

Fourier transform of a sawtooth function yields a spectrum with spectral components at harmonics of the oscillation frequency decreasing as $1/n^2$. These harmonics are visible in the emission spectrum displayed in Fig. \ref{EmissionStrong}, the measurement conditions of which corresponds to the time trace in Fig. \ref{Nr(t)}. The ratios of peak areas amount to 0.30 and 0.18 for the second and third harmonic against the first one, respectively. For a sawtooth function these ratios are 1:4 and 1:9, respectively.  The agreement between these first few harmonics is good, which supports our model for the  time dependence of the cavity photon occupation as regulated by the thermal self-oscillations.

\section{Conclusion}
To conclude, we have studied superconducting microwave cavity coupled to a monolayer graphene flake. The graphene flake acts as a SINIS resistor and has a strong nonlinear dependence with temperature. This nonlinearity leads to thermal runaway at a range of temperatures where the electron-phonon coupling is not able to balance the increased heating induced by $V^2/R_{\mathrm{SINIS}}(T)$ while driving the cavity to higher occupation number of photons. Owing to large dissipation in the normal state, self-oscillations appear and the cavity switches between high and low $Q$ states in the oscillation. The self-oscillations are seen in emission measurements as a sequence of sidepeaks, the magnitudes of which indicate nearly sawtooth-like time dependence for the number of quanta in the cavity. 
Cavity transmission measurements were employed to characterize the thermal oscillation sideband frequency $f_{\mathrm{SB}}$ as a function of temperature and pumping power. 
The sideband frequency $f_{\mathrm{SB}}$ increases with applied pump power almost linearly 
while with temperature, $f_{\mathrm{SB}}$ shows $T^3$ dependence below 500mK. We employ a thermal oscillation model to simulate the behavior and extract the electron-phonon coupling constant $\Sigma=2.2$\,W/m$^2$/K$^3$ from the data. The obtained value for $\Sigma$ agrees well with theoretical heat flow estimates for disordered graphene at $T < T_{\mathrm{dis}}$ using $D=50$\,eV and typical graphene to rough metal contact properties.

\section*{Acknowledgments}

We are grateful to M. Dykman, S. H. Raja, H. Sepp\"a, and V. Shumeiko for useful discussions. This work was supported by the Academy of Finland
  projects 314448 (BOLOSE) and 336813 (CoE, Quantum Technology
  Finland) as well as by ERC (grant no. 670743). The research leading to these results has received funding from the European Union’s Horizon 
2020 Research and Innovation Programme, under Grant Agreement no 824109, and the experimental work benefited from
  the Aalto University OtaNano/LTL infrastructure. MTH  acknowledges support from the European Union’s Horizon  2020  Programme  for  Research  and  Innovation under   grant   agreement   No.722923   (Marie   Curie ETN   -   OMT).


\appendix
\section{Modelling of the system} \label{a1}
\begin{center}
\begin{figure}[htb]
\includegraphics[width=\linewidth]{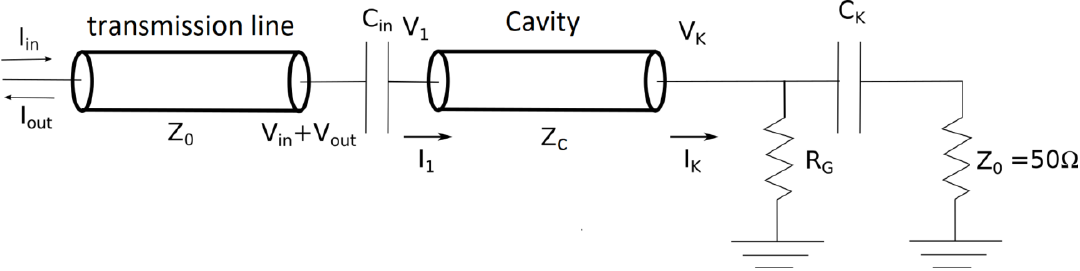}
\caption{Graphene junction with resistance $R_G$ connected to a resonator with the characteristic impedance $Z_C\approx 100\,\Omega$. 
Resistance of the graphene flake is $R_G=R_{\mathrm{SINIS}}(T)$ for $T<T_C$ and $R_G=R_N\approx 0.5$ k$\Omega$ for $T>T_C$, $T_C=9$ K is the critical
temperature of the superconducting MoRe leads,
the input capacitor $C_{\rm in}=0.5$ fF and the coupling capacitor to external readout device with $Z_0$ is $C_K=2.21$ fF.}
\label{system}
\end{figure}
\end{center}

We consider the system depicted in Fig. \ref{system}. It consists 
of the superconductor - insulator - normal graphene - insulator - superconductor (SINIS) junction coupled to a resonator.
The Josephson effect in this SINIS junction is absent, and it acts as an effective resistor. 
We assume that the resistance of the SINIS structure is due to quasiparticle transport via the insulating barriers and
it depends on the temperature as follows
\begin{equation}
\frac{1}{R_{\mathrm{SINIS}}(T)} = \frac{1}{R_0} +  \frac{1}{R_N}\exp\left[ -\frac{\Delta}{k_BT} \right],
\label{Rqp}
\end{equation}
where $R_0$ is the subgap resistance of the cavity,
$R_N$ is the resistance of the structure in the normal state and $\Delta$ is the superconducting gap in the MoRe leads.
In the expriment one finds $R_N\approx 0.5$ k$\Omega$, $T_C=9$K, $\Delta=1.76 k_BT_C = 2.187\times 10^{-22}$ J ($\Delta=1.37$ meV), and $R_0=3 \times 10^6$\,$\Omega$. For the most part, $R_0$ can be considered a constant, although hot quasiparticles are found to influence its value at large drives. We take this into account by using an alternate subgap resistance value $R_0^* = R_{00}$exp$(\Delta/k_B \sqrt{T_0^2 + (\alpha T)^2})$ where $R_{00} \sim 10$\,$\Omega$ and $\alpha $ is a tunable parameter $0 < \alpha < 1 $.

If $T<T_C$, the electron temperature of the graphene flake obeys the equation
\begin{equation}
C_G(T)\frac{dT}{dt} = -\Sigma A (T^\gamma - T_0^\gamma) + P_{\rm diss}.
\label{heat}
\end{equation}
Here $C_G$ is the electronic heat capacity of graphene, $\Sigma$ is the material constant describing cooling of the
graphene electrons via phonons, $A$ is the area of the flake,
$T_0$ is the bath temperature and $P_{\rm diss}$ is the part of the incoming microwave pumping power $P_{\rm in}$,  
which penetrates through the resonator and is absorbed in graphene.
The heat capacity of graphene is known \cite{Aamir2021}
\begin{equation}
C_G(T) = \frac{2Ak_B}{\pi\hbar^2v_F^2}\left[ \frac{\pi^2}{3} |\mu| k_B T + \frac{9\zeta(3)}{2}k_B^2T^2 \right].
\label{Ce}
\end{equation}
Here $v_F=10^6$ m/s is the Fermi velocity in graphene and $\mu$ is the chemical potential tunable by gate voltage.
Eq. (\ref{Ce}) has been verified in the experiment \cite{Aamir2021}.
The parameters $\Sigma$ and $\gamma$ in Eq. (\ref{heat}) are expected to be close to those in our recent paper \cite{Tomi2021}.
There we have found $\Sigma A\approx 5$ nW$/T_C^\gamma$ and $\gamma=3.1$. However, these parameters may take different values 
because Al leads are not the same as MoRe ones, and because the size of the flake is different. Consequently, we set $\gamma=3$ and regard $\Sigma$ as a fit parameter.

The potential $V_K$ inside resonator in the vicinity of the coupling capacitor $C_K$ (see Fig. \ref{system}) obeys the equation  
\begin{eqnarray}
\ddot V_K + \kappa_T(T)\dot V_K + \omega_r^2 V_K 
= \frac{2\kappa_{\rm in}}{C_{\rm in}} I_{\rm in} + \frac{2Z_C\omega_r^2}{\pi} \xi_G(t),\hspace{0.7cm}
\label{eq}
\end{eqnarray}
where $\kappa_T(T) = \kappa_{\rm in} + \kappa_{\rm out} +\kappa_G(T)$ is the sum of the damping rates of the resonator due to the coupling to the input transmission line, to the cavity readout impedance $Z_0$ at the output terminal,
and to the SINIS structure:
\begin{eqnarray}
\kappa_{\rm in} = \frac{2\omega_r^3 Z_C Z_0 C_{\rm in}^2}{\pi}, \\
\kappa_{\rm out} = \frac{2 Z_C Z_0\omega_r^3C_K^2}{\pi}, \\
\kappa_G(T) = \frac{2Z_C\omega_r}{\pi R_G(T)}.\,\,\,\,
\end{eqnarray}
Further, $\xi_G(t)$ is the Nyquist noise of SINIS with the spectral density
\begin{equation}
\langle |\xi_G|^2_\omega \rangle  = \frac{\hbar\omega}{R_G(T)}\coth\frac{\hbar\omega}{2k_BT}.
\end{equation}

\subsection{Thermal oscillations}

Since we are interested in slow thermal relaxation oscillations, we neglect the effects due to electron-electron interactions \cite{Voutilainen2011}, and re-write Eq. (\ref{eq}) in terms of
a slow variable --- the number of photons in the resonator $N_r$. This parameter is defined in terms of the energy of the resonator $E_r$,
\begin{eqnarray}
N_r = \frac{E_r}{\hbar\omega_r}  = \frac{\pi ( \dot V_K^2 + \omega_r^2 V_K^2 )}{4 Z_C\hbar\omega_r^4}.
\end{eqnarray}
Solving Eq. (\ref{eq}), 
we find the steady state number of photons $N_{st}$ in presence of the
sinusoidal pump signal of the form $I_{\rm in}(t)=I_p\sin\omega_pt$,
\begin{widetext}
\begin{equation}
N_{st} = \frac{\kappa_{\rm in}}{\kappa_T(T)}\frac{1}{2}\coth\frac{\hbar\omega_r}{2k_BT_0}
+\frac{8 Z_C P_{\rm in}}{\hbar Z_0} \frac{\kappa_{\rm in}\omega_r}{(\omega_p^2-\omega_r^2)^2+\omega_p^2 \kappa_T(T) ^2}
+ \frac{\kappa_G(T)}{\kappa_T(T)}\frac{1}{2}\coth\frac{\hbar\omega_r}{2k_BT},
\end{equation}
\end{widetext}
where  $P_{\rm in}=I_p^2 Z_0/2$ is the incoming pumping power and $\kappa_T(T)=\kappa_{\rm in} + \kappa_{\rm out} +\kappa_G(T)$.
The first term in this expression is due to the thermal radiation coming from the input capacitor, the last term is
caused by the Nyquist noise of the SINIS structure, and the middle term is the non-equilibrium population of the resonator 
induced by the pumping sinusoidal signal.
We will consider strong pumping regime, where we can approximate
\begin{eqnarray}
N_{st}(T) = \frac{8 Z_C P_{\rm in}}{\hbar Z_0} \frac{\kappa_{\rm
in}\omega_r}{(\omega_p^2-\omega_r^2)^2+\omega_p^2 \kappa_T(T)^2}.
\label{Nst}
\end{eqnarray}
The steady state value $N_{st}(T)$ depends on temperature via the damping rate  $\kappa_G(T)$ of graphene  in $\kappa_T(T)$.

With these preparations, Eqs. (\ref{eq}) and (\ref{heat}) can be written in the form
\begin{eqnarray}
\frac{dN_r}{dt} &=& -\kappa_T(T)(N_r-N_{st}(T)) + \zeta(t),
\nonumber\\
C_G(T)\frac{dT}{dt} &=& - \Sigma A(T^\gamma - T_0^\gamma) + \hbar\omega_r \kappa_G(T) N_r.
\label{dNdt}
\end{eqnarray}
Here we have made the following approximations: the dissipated power appearing in Eq. (\ref{heat}) is expressed as $P_{\rm diss}=\hbar\omega_r \kappa_G(T) N_r$, 
and $\zeta(t)$ is the noise term given by the equation
\begin{eqnarray}
\zeta(t) = \frac{\dot V_k\xi_G(t)}{\hbar\omega_r^2}.
\end{eqnarray} 
Eqs. (\ref{dNdt}) can be easily solved numerically if we put $\zeta(t)=0$.

To understand the origin of thermal oscillations, we assume that the heat capacity of graphene is very small and put $C_G(T)=0$.
This assumption should reasonably well correspond to the experimental situation. After that, 
from the second equation (\ref{dNdt}) we obtain
\begin{eqnarray}
N_r(T) = \frac{\Sigma A(T^\gamma -T_0^\gamma)}{\hbar\omega_r \kappa_G(T)}.
\label{NrT}
\end{eqnarray}
Substituting the result in the first equation, we obtain single equation for the temperature, which describes the system,
\begin{eqnarray}
\frac{dN_r(T)}{dT} \frac{dT}{dt} = -\kappa_T(T)( N_r(T) - N_{st}(T)).
\label{eq1}
\end{eqnarray}

\begin{center}
\begin{figure}[htb]
\includegraphics[width=0.9\linewidth]{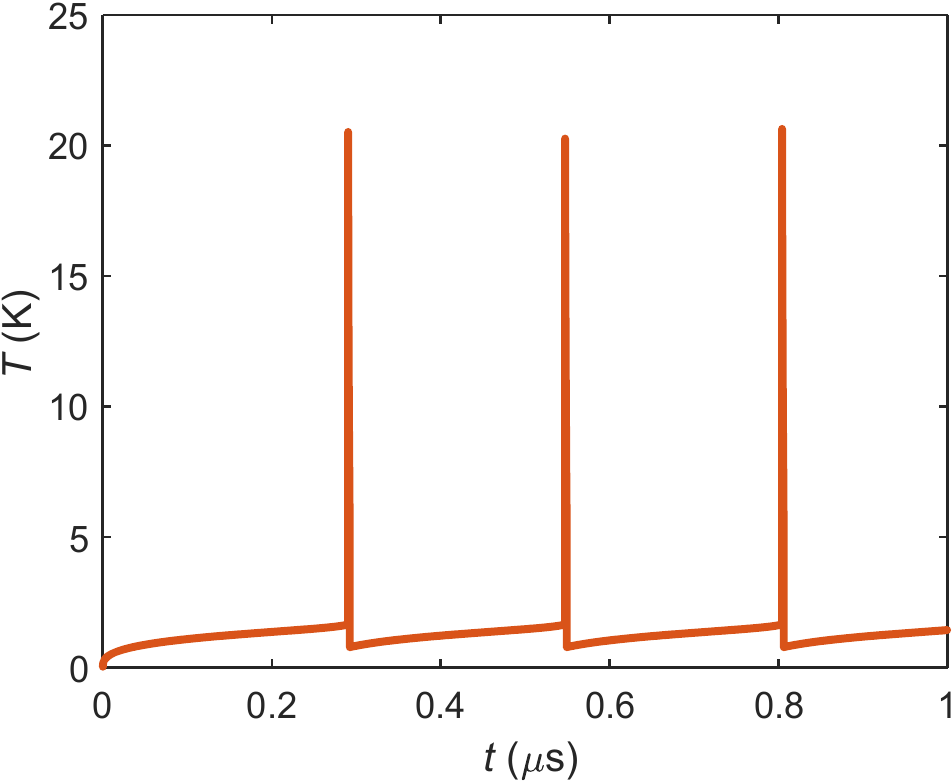}
\caption{Time dependence of the temperature, $T(t)$, obtained by numerically solving the equations (\ref{dNdt}).
The system parameters are the same as in Fig. \ref{Fig:Nr_T}. In addition, we have set the drive power $P=-53$ dBm and
the chemical potential $\mu=10$ meV. }
\label{T(t)}
\end{figure}
\end{center}

For certain values of the system parameters the dependence $N_r(T)$, given by Eq. (\ref{NrT}), is non-monotonous. 
In Fig. \ref{Fig:Nr_T} we plot this dependence for certain choice of the 
parameters indicated in the figure caption. The function $N_r(T)$ reaches maximum at temperature $T_1$ and minimum at temperature $T_2$. We also define two other temperatures: $T_{\min}$ and $T_{\max}$ as indicated in Fig. \ref{Fig:Nr_T}.
The temperature of the steady state $T_{\rm st}$ for a given power $P_{\rm in}$ can be found from Eqs. (\ref{dNdt})
(or by equalizing the right-hand side of Eq. (\ref{eq1}) to zero), 
and it is the solution of the equation
\begin{eqnarray}
\Sigma A(T_{\rm st}^\gamma - T_0^\gamma) = \hbar\omega_r \kappa_G(T_{\rm st}) N_{\rm st}(T_{\rm st}).
\label{steady}
\end{eqnarray}
The temperature $T_{\rm st}$ grows with the applied power $P_{\rm in}$. At small powers, $P_{\rm in}<P_{\min}$, this temperature
stays below $T_1$, 
$T_{\rm st}<T_1$, the steady state is stable and no temperature oscillations occur.  
At intermediate powers, $P_{\min} < P_{\rm in} < P_{\max}$, the solution of Eq. (\ref{steady}) shifts to the interval $T_{1}<T_{\rm st}<T_{2}$,
where the steady state becomes unstable due to the negative sign of the derivative $dN_r(T)/dT$ in the left-hand side of Eq. (\ref{eq1}). 
In this case, the temperature periodically changes along the circle indicated by red arrows in Fig. \ref{Fig:Nr_T}.
The time dependence of the temperature of graphene, $T(t)$, for this regime is shown in Fig. \ref{T(t)}, 
and the time dependence of the number of photons in the resonator $N_r(t)$ --- in Fig. \ref{Nr(t)}. 
Finally, at sufficiently high powers, $P_{\rm in}>P_{\max}$, the solution of Eq. (\ref{steady}) exceeds the temperature $T_{\max}$,
i.e. $T_{\rm st}>T_{2}$, the steady state again becomes stable and the thermal oscillations disappear.

\section{Supporting Experimental Results} \label{SuppExp}
\subsection{Conductance and mobility}
\begin{center}
\begin{figure}[htb]
\includegraphics[width=0.95\linewidth]{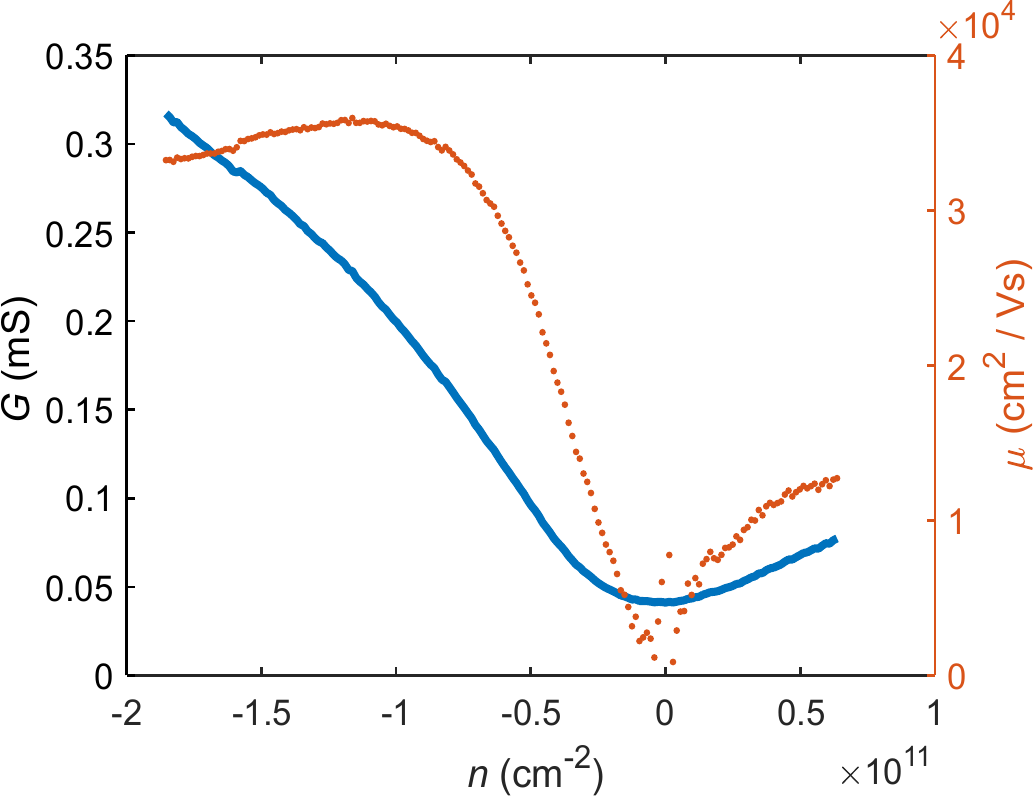}
\caption{DC conductance $G$ of the sample (blue solid trace) as a function of carrier density $n=C_g(V_g-V_g^D)/e$ measured at 4\,K; the Dirac point is located at $V_g^D = 1.7$ V. The red points indicate the mobility (right scale) evaluated from the conductance (see text).}
\label{conductance_charge_density}
\end{figure}
\end{center}

Fig. \ref{conductance_charge_density} displays measured conductance at $T_0=4$ K. The  graphene conductance at Dirac point $G_G(n_0)$ is reduced here by the double SIG interfacial resistance of the sample: $G_{\mathrm{min}}=\left[1/G_G(n_0)+2R_c\exp(\Delta/k_BT)\right]^{-1}$ where $R_c\simeq 200 $ $\Omega$ is the normal state resistance of a single SIG interface. The variation of the graphene with charge density $n$ is clearly visible in Fig. \ref{conductance_charge_density} around $n=1\times10^{11}$ cm$^{-2}$. Consequently, the slope of the data $dG/dn$ can be employed for determination of mobility using $\mu =\frac{W}{L} (G-G_{\mathrm{min}})/n e$. By accounting for the enhanced contact resistance $R_c$ in the superconducting state, we obtain for the mobility $\mu \sim 3.5$ m$^{2}$/Vs. The mean free path $\ell$ was obtained from the semiclassical formula for conductivity $\sigma=\frac{2e^2}{\hbar} \sqrt{\pi n_g} \ell$. The value of contact resistance at 4\,K is estimated using the normal state resistance, and assuming that the interfacial resistance $R_{\mathrm{SIG}}$ follows the gate induced charge, we obtain a mean free path of 100 nm for electrons in the suspended graphene section.

\subsection{Kinetic inductance of the cavity}
Although the center conductor in the MoRe cavity has a rather large cross section 3\,$\mu$m$^2$, the carrier concentration in it is rather small, as can be deduced from the normal state resistance of the cavity $\sim 1$ k$\Omega$. Consequently, the kinetic energy of Cooper pairs will contribute to effective inductance of the cavity. As the density of Cooper pairs decreases with $T$, kinetic energy grows and the effective inductance increases, which decrease the cavity frequency. The decrease in $f$ up to $T=2$ \,K is on the order of the cavity line width, which allows quite accurate determination of the electron temperature in the thermal oscillation regime around $T =1.5 - 2$\,K.  
\begin{center}
\begin{figure}[htb]
\centering
\includegraphics[width=0.99\linewidth]{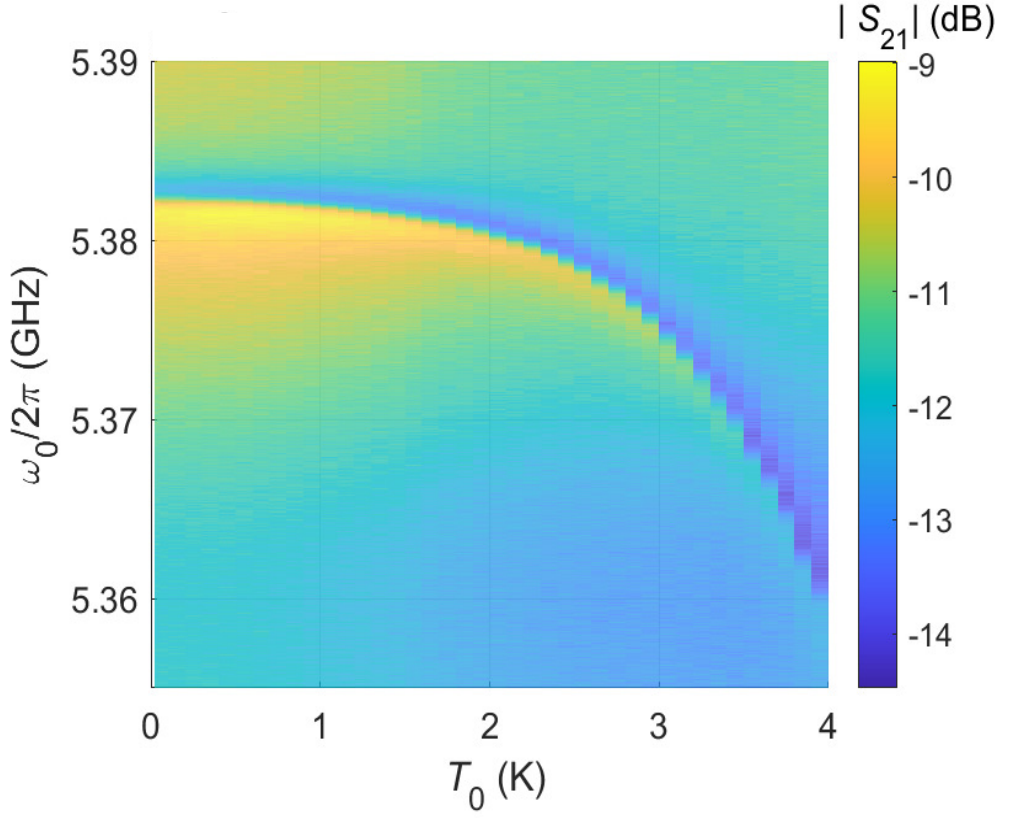}
\caption{Cavity resonance frequency $f_0=\omega_0/2\pi$ as a function of cryostat temperature $T_0$. The shift in the resonance position is due to decrease in the kinetic inductance of MoRe superconductor towards higher temperatures.}
\label{cavity_vs_temp}
\end{figure}
\end{center}

\subsection{Fitting of the resonance line shape}
The measured transmission signal $S_{21}(f)$ displays Fano-resonance character (cf. Fig. \ref{cavity_vs_power}), which is typical for  $\lambda/2$ microwave cavity transmission  with parasitic capacitive shunting. The shape of a Fano-resonance can be parametrized as
\begin{equation}
    S_{21}^F(\omega)=\frac{\left(q\Gamma/2+\omega-\omega_0 \right)^2}{\left(\Gamma/2\right)^2+\left(\omega-\omega_0\right)^2},
    \label{Cavity_fiTquation}
\end{equation}
where $q$ is the Fano parameter, $\Gamma$ is the resonance width (decay rate) and $\omega-\omega_0$ is the probe frequency minus the resonance frequency. A typical measured resonance is shown in Fig. \ref{cavity_vs_power} by the trace at  $P_{\mathrm{in}}=-62$\,dBm, which fits well Eq. (\ref{Cavity_fiTquation}) using $q=-1.08$. 

\begin{center}
\begin{figure}[htb]
\centering
\includegraphics[width=0.99\linewidth]{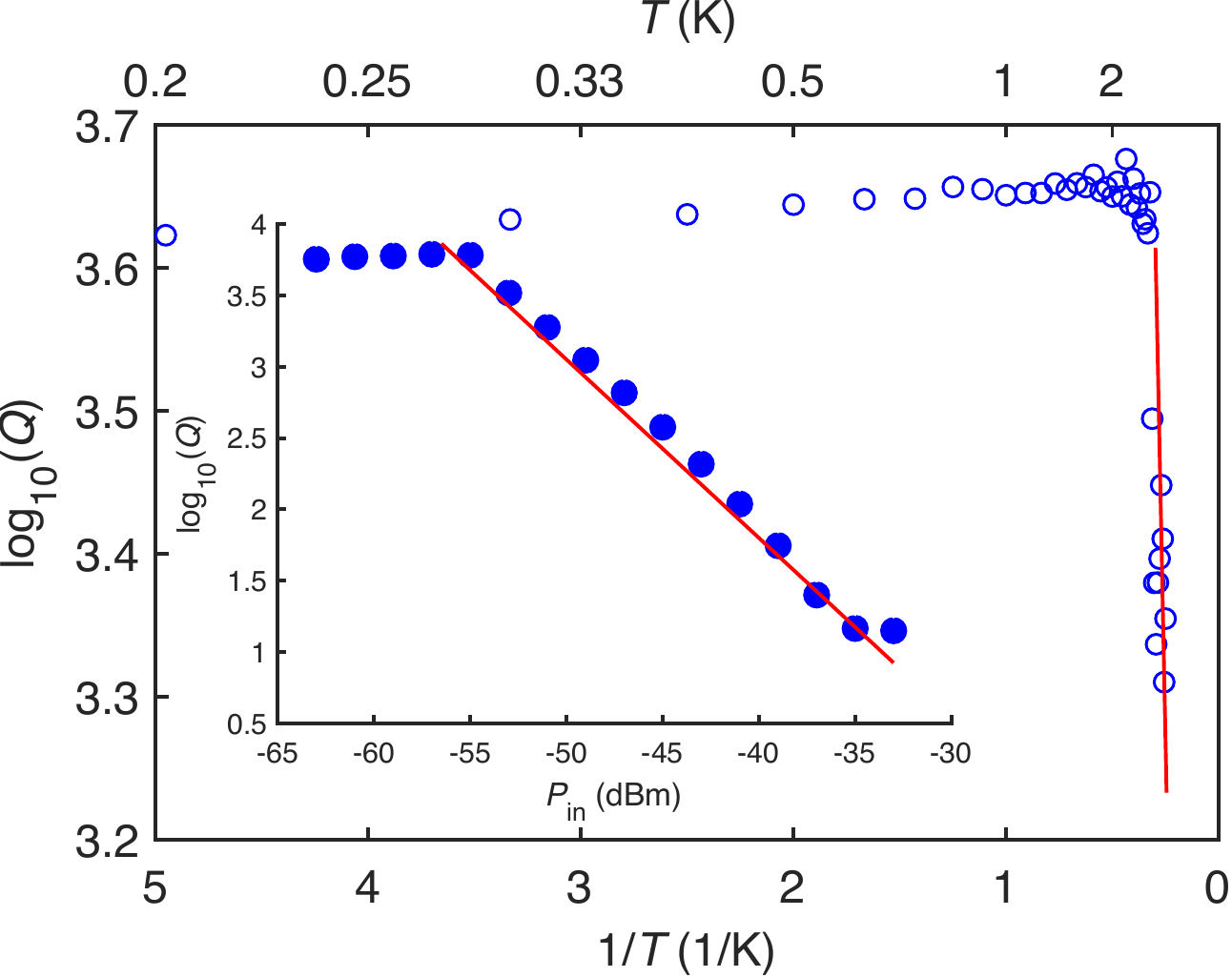}
\caption{Logarithm of the $Q$-factor as a function of the inverse temperature $1/T$. The red line indicates a slope of $\Delta/k_B=15$\,K. Points measured at $T<0.2$\,K are not shown, but they are in line with the trend of the other points. Inset: Quality factor $Q$ as a function of the input power $P_{\mathrm{in}}$ injected at resonance frequency $f(P_{\mathrm{in}})$. The overlaid line displays a slope of $-1/3$, which on a log-log scale indicates power law dependence $Q \propto P_{\mathrm{in}}^{1/3}$.}
\label{resonance_vs_temperature}
\end{figure}
\end{center}

For fitting the temperature dependent resonances $q$, $\Gamma$ and $\omega_0$ were adjustable parameters, with $q$ ranging from $-1.05$ to $-1.47$. The obtained $Q=\frac{\omega_0}{\Gamma}$ is displayed in Fig. \ref{resonance_vs_temperature} as a function of inverse temperature $1/T$. At $T>2$\,K, the data follow activation type of behavior $Q(T) \propto \exp(\Delta/k_BT)$ with $\Delta/k_B \simeq 15$\,K. The $Q$ value determination based on  Eq. (\ref{Cavity_fiTquation}) was verified by circuit impedance analysis using a cavity with a capacitive shunt, and good agreement for the line shape and width were obtained by the fitted impedance parameters of this circuit model. From the circuit impedance analysis the capacitive shunt is estimated by $C_\mathrm{parasitic} \approx 4.75$\,fF.

\subsection{Reduction of quality factor with drive power}

The transmission signals are more difficult to analyze at higher powers as the thermal oscillation makes a step wise change in the transmission at a frequency dependent on the applied power. To fit the traces at higher input power, it is assumed that only the central part of the Fano resonance is visible. At the Fano resonance's center, the slope $a=\frac{4q}{\Gamma}$ and the shape can be approximated by a linear function. However, to extract $\Gamma$ from this equation $q$ has to be determined separately. For all transmission signals $S_{21}(f)$ in the self-oscillation regime, we used the same value $q=-1.08$ obtained from the undisturbed resonance fit. Even when cross-comparing the value $q=-1.08$ with the 
$q$ values obtained from equilibrium transmission data at different temperatures, the $Q$-factor changes by orders of magnitude (see inset Fig. \ref{resonance_vs_temperature}), whereas $q$ only varies by a factor on the order of $1.5$. Hence, the error in $Q$ factor determination can be regarded as small.

The data in the inset of Fig. \ref{resonance_vs_temperature} indicate a reduction of $\log Q \propto P_{\mathrm{in}}^{-1/3}$ at drive powers above -55\,dBm across the whole range of powers used in our thermal oscillation measurements. Neglecting the threshold in power, we may equate this dependence by the temperature dependence $\log Q \propto \Delta/k_BT$. Consequently, we obtain a relation  $P_{\mathrm{in}} \propto T^3$ for temperatures in the thermal self-oscillation regime. This agrees exactly with the supposed electron-phonon coupling interaction with $\gamma=3$ in our graphene.

\subsection{Determination of the threshold drive power}

\begin{center}
\begin{figure}[htb]
\centering
\includegraphics[width=0.95\linewidth]{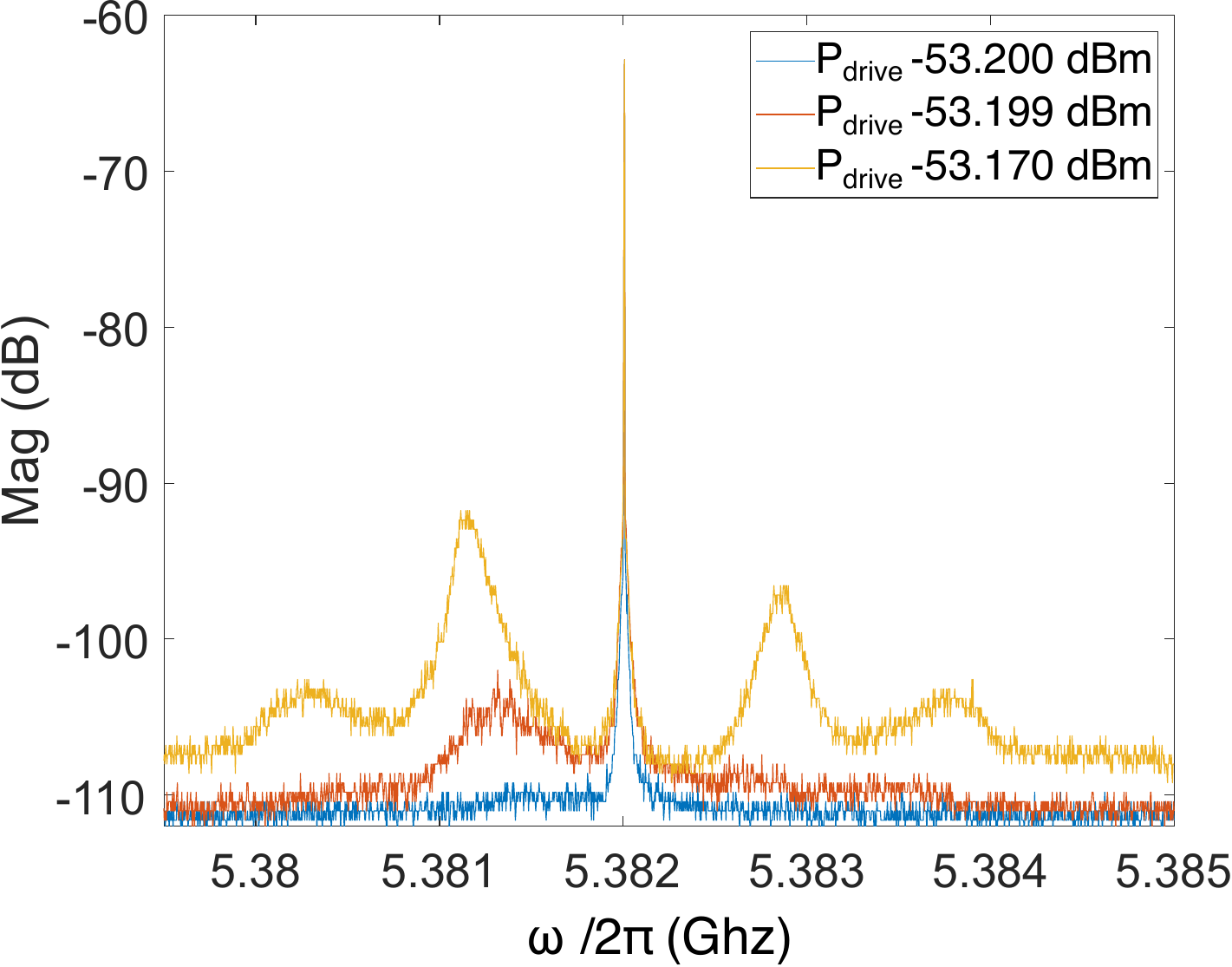}
\caption{Emission spectrum of the thermal oscillations when the cavity is pumped at resonance above threshold power. Thermal sideband peaks shift with pump power. The blue trace at $P_{\mathrm{in}}=-53.200\mathrm{ dBm }=P_{\mathrm{th}}$ defines the threshold power; $P_{\mathrm{in}}-P_{\mathrm{th}}= 0.001$ dBm yields a clear side peak as seen in the red trace. }
\label{ThermalOscillationSpectrum}
\end{figure}
\end{center}

The onset of thermal oscillation requires sufficient drive, the strength of which depends on the detuning from the cavity resonance. To determine the smallest threshold power for thermal oscillation, we have varied the pumping frequency around the ground state cavity resonance frequency and measured the emission of the pumped circuit. Fig. \ref{ThermalOscillationSpectrum} displays the emission spectra obtained at the smallest pumping powers. The threshold power is identified as $P_{\mathrm{th}}=-53.200$ dBm; at $P_{\mathrm{in}}=-53.199$ dBm a clear side peak for a thermal oscillation at $f_{\mathrm{SB}} \simeq 0.7$ MHz is seen. Note also that the half width of the side peak is about 0.5 MHz, which presumably indicates strong variation in the switching times owing to fluctuations in the number of quanta in the cavity.

\twocolumngrid
\bibliography{Collection}

\begin{thebibliography}{41}%
\makeatletter
\providecommand \@ifxundefined [1]{%
 \@ifx{#1\undefined}
}%
\providecommand \@ifnum [1]{%
 \ifnum #1\expandafter \@firstoftwo
 \else \expandafter \@secondoftwo
 \fi
}%
\providecommand \@ifx [1]{%
 \ifx #1\expandafter \@firstoftwo
 \else \expandafter \@secondoftwo
 \fi
}%
\providecommand \natexlab [1]{#1}%
\providecommand \enquote  [1]{``#1''}%
\providecommand \bibnamefont  [1]{#1}%
\providecommand \bibfnamefont [1]{#1}%
\providecommand \citenamefont [1]{#1}%
\providecommand \href@noop [0]{\@secondoftwo}%
\providecommand \href [0]{\begingroup \@sanitize@url \@href}%
\providecommand \@href[1]{\@@startlink{#1}\@@href}%
\providecommand \@@href[1]{\endgroup#1\@@endlink}%
\providecommand \@sanitize@url [0]{\catcode `\\12\catcode `\$12\catcode
  `\&12\catcode `\#12\catcode `\^12\catcode `\_12\catcode `\%12\relax}%
\providecommand \@@startlink[1]{}%
\providecommand \@@endlink[0]{}%
\providecommand \url  [0]{\begingroup\@sanitize@url \@url }%
\providecommand \@url [1]{\endgroup\@href {#1}{\urlprefix }}%
\providecommand \urlprefix  [0]{URL }%
\providecommand \Eprint [0]{\href }%
\providecommand \doibase [0]{http://dx.doi.org/}%
\providecommand \selectlanguage [0]{\@gobble}%
\providecommand \bibinfo  [0]{\@secondoftwo}%
\providecommand \bibfield  [0]{\@secondoftwo}%
\providecommand \translation [1]{[#1]}%
\providecommand \BibitemOpen [0]{}%
\providecommand \bibitemStop [0]{}%
\providecommand \bibitemNoStop [0]{.\EOS\space}%
\providecommand \EOS [0]{\spacefactor3000\relax}%
\providecommand \BibitemShut  [1]{\csname bibitem#1\endcsname}%
\let\auto@bib@innerbib\@empty
\bibitem [{\citenamefont {Jenkins}(2013)}]{Jenkins2013}%
  \BibitemOpen
  \bibfield  {author} {\bibinfo {author} {\bibfnamefont {A.}~\bibnamefont
  {Jenkins}},\ }\href {\doibase https://doi.org/10.1016/j.physrep.2012.10.007}
  {\bibfield  {journal} {\bibinfo  {journal} {Physics Reports}\ }\textbf
  {\bibinfo {volume} {525}},\ \bibinfo {pages} {167} (\bibinfo {year}
  {2013})}\BibitemShut {NoStop}%
\bibitem [{\citenamefont {Betz}\ \emph
  {et~al.}(2012{\natexlab{a}})\citenamefont {Betz}, \citenamefont {Jhang},
  \citenamefont {Pallecchi}, \citenamefont {Ferreira}, \citenamefont
  {F{\`{e}}ve}, \citenamefont {Berroir},\ and\ \citenamefont
  {Pla{\c{c}}ais}}]{Betz2012}%
  \BibitemOpen
  \bibfield  {author} {\bibinfo {author} {\bibfnamefont {A.~C.}\ \bibnamefont
  {Betz}}, \bibinfo {author} {\bibfnamefont {S.~H.}\ \bibnamefont {Jhang}},
  \bibinfo {author} {\bibfnamefont {E.}~\bibnamefont {Pallecchi}}, \bibinfo
  {author} {\bibfnamefont {R.}~\bibnamefont {Ferreira}}, \bibinfo {author}
  {\bibfnamefont {G.}~\bibnamefont {F{\`{e}}ve}}, \bibinfo {author}
  {\bibfnamefont {J.-M.}\ \bibnamefont {Berroir}}, \ and\ \bibinfo {author}
  {\bibfnamefont {B.}~\bibnamefont {Pla{\c{c}}ais}},\ }\href {\doibase
  10.1038/nphys2494} {\bibfield  {journal} {\bibinfo  {journal} {Nature
  Physics}\ }\textbf {\bibinfo {volume} {9}},\ \bibinfo {pages} {109} (\bibinfo
  {year} {2012}{\natexlab{a}})}\BibitemShut {NoStop}%
\bibitem [{\citenamefont {Laitinen}\ \emph {et~al.}(2014)\citenamefont
  {Laitinen}, \citenamefont {Oksanen}, \citenamefont {Fay}, \citenamefont
  {Cox}, \citenamefont {Tomi}, \citenamefont {Virtanen},\ and\ \citenamefont
  {Hakonen}}]{Laitinen2014}%
  \BibitemOpen
  \bibfield  {author} {\bibinfo {author} {\bibfnamefont {A.}~\bibnamefont
  {Laitinen}}, \bibinfo {author} {\bibfnamefont {M.}~\bibnamefont {Oksanen}},
  \bibinfo {author} {\bibfnamefont {A.}~\bibnamefont {Fay}}, \bibinfo {author}
  {\bibfnamefont {D.}~\bibnamefont {Cox}}, \bibinfo {author} {\bibfnamefont
  {M.}~\bibnamefont {Tomi}}, \bibinfo {author} {\bibfnamefont {P.}~\bibnamefont
  {Virtanen}}, \ and\ \bibinfo {author} {\bibfnamefont {P.~J.}\ \bibnamefont
  {Hakonen}},\ }\href {\doibase 10.1021/nl404258a} {\bibfield  {journal}
  {\bibinfo  {journal} {Nano Letters}\ }\textbf {\bibinfo {volume} {14}},\
  \bibinfo {pages} {3009} (\bibinfo {year} {2014})}\BibitemShut {NoStop}%
\bibitem [{\citenamefont {Song}\ \emph {et~al.}(2012)\citenamefont {Song},
  \citenamefont {Reizer},\ and\ \citenamefont {Levitov}}]{Song2012}%
  \BibitemOpen
  \bibfield  {author} {\bibinfo {author} {\bibfnamefont {J.~C.~W.}\
  \bibnamefont {Song}}, \bibinfo {author} {\bibfnamefont {M.~Y.}\ \bibnamefont
  {Reizer}}, \ and\ \bibinfo {author} {\bibfnamefont {L.~S.}\ \bibnamefont
  {Levitov}},\ }\href {\doibase 10.1103/PhysRevLett.109.106602} {\bibfield
  {journal} {\bibinfo  {journal} {Physical Review Letters}\ }\textbf {\bibinfo
  {volume} {109}},\ \bibinfo {pages} {106602} (\bibinfo {year} {2012})},\
  \Eprint {http://arxiv.org/abs/1111.4678} {arXiv:1111.4678} \BibitemShut
  {NoStop}%
\bibitem [{\citenamefont {Chen}\ and\ \citenamefont {Clerk}(2012)}]{Clerk2012}%
  \BibitemOpen
  \bibfield  {author} {\bibinfo {author} {\bibfnamefont {W.}~\bibnamefont
  {Chen}}\ and\ \bibinfo {author} {\bibfnamefont {A.~A.}\ \bibnamefont
  {Clerk}},\ }\href {\doibase 10.1103/PhysRevB.86.125443} {\bibfield  {journal}
  {\bibinfo  {journal} {Physical Review B}\ }\textbf {\bibinfo {volume} {86}},\
  \bibinfo {pages} {125443} (\bibinfo {year} {2012})}\BibitemShut {NoStop}%
\bibitem [{\citenamefont {Kubakaddi}(2009)}]{Kubakaddi2009}%
  \BibitemOpen
  \bibfield  {author} {\bibinfo {author} {\bibfnamefont {S.~S.}\ \bibnamefont
  {Kubakaddi}},\ }\href {\doibase 10.1103/PhysRevB.79.075417} {\bibfield
  {journal} {\bibinfo  {journal} {Phys. Rev. B}\ }\textbf {\bibinfo {volume}
  {79}},\ \bibinfo {pages} {75417} (\bibinfo {year} {2009})}\BibitemShut
  {NoStop}%
\bibitem [{\citenamefont {Tse}\ and\ \citenamefont {{Das
  Sarma}}(2009)}]{Tse2009}%
  \BibitemOpen
  \bibfield  {author} {\bibinfo {author} {\bibfnamefont {W.-K.}\ \bibnamefont
  {Tse}}\ and\ \bibinfo {author} {\bibfnamefont {S.}~\bibnamefont {{Das
  Sarma}}},\ }\href {\doibase 10.1103/PhysRevB.79.235406} {\bibfield  {journal}
  {\bibinfo  {journal} {Physical Review B}\ }\textbf {\bibinfo {volume} {79}},\
  \bibinfo {pages} {235406} (\bibinfo {year} {2009})}\BibitemShut {NoStop}%
\bibitem [{\citenamefont {Bistritzer}\ and\ \citenamefont
  {MacDonald}(2009)}]{Bistritzer2009}%
  \BibitemOpen
  \bibfield  {author} {\bibinfo {author} {\bibfnamefont {R.}~\bibnamefont
  {Bistritzer}}\ and\ \bibinfo {author} {\bibfnamefont {A.}~\bibnamefont
  {MacDonald}},\ }\href {\doibase 10.1103/PhysRevLett.102.206410} {\bibfield
  {journal} {\bibinfo  {journal} {Physical Review Letters}\ }\textbf {\bibinfo
  {volume} {102}},\ \bibinfo {pages} {206410} (\bibinfo {year}
  {2009})}\BibitemShut {NoStop}%
\bibitem [{\citenamefont {Viljas}\ and\ \citenamefont
  {Heikkil\"a}(2010)}]{Viljas2010}%
  \BibitemOpen
  \bibfield  {author} {\bibinfo {author} {\bibfnamefont {J.~K.}\ \bibnamefont
  {Viljas}}\ and\ \bibinfo {author} {\bibfnamefont {T.~T.}\ \bibnamefont
  {Heikkil\"a}},\ }\href {http://dx.doi.org/10.1103/PhysRevB.81.245404}
  {\bibfield  {journal} {\bibinfo  {journal} {Phys. Rev. B}\ }\textbf {\bibinfo
  {volume} {81}},\ \bibinfo {pages} {245404} (\bibinfo {year}
  {2010})}\BibitemShut {NoStop}%
\bibitem [{\citenamefont {Suzuura}\ and\ \citenamefont
  {Ando}(2002)}]{Suzuura2002}%
  \BibitemOpen
  \bibfield  {author} {\bibinfo {author} {\bibfnamefont {H.}~\bibnamefont
  {Suzuura}}\ and\ \bibinfo {author} {\bibfnamefont {T.}~\bibnamefont {Ando}},\
  }\href {\doibase 10.1103/PhysRevB.65.235412} {\bibfield  {journal} {\bibinfo
  {journal} {Physical Review B}\ }\textbf {\bibinfo {volume} {65}},\ \bibinfo
  {pages} {235412} (\bibinfo {year} {2002})}\BibitemShut {NoStop}%
\bibitem [{\citenamefont {Katsnelson}(2012)}]{Katsnelson2012}%
  \BibitemOpen
  \bibfield  {author} {\bibinfo {author} {\bibfnamefont {M.~I.}\ \bibnamefont
  {Katsnelson}},\ }\href
  {http://books.google.com/books?hl=en&lr=&id=FgwW3aVRBT0C&oi=fnd&pg=PR7&dq=graphene+carbon+in+two+dimensions&ots=DfomUbm4eh&sig=YJXxO9t5cdXV6Ehki5Lf_UAq2dQ}
  {\emph {\bibinfo {title} {{Graphene: Carbon in Two Dimensions}}}},\ \bibinfo
  {edition} {1st}\ ed.\ (\bibinfo  {publisher} {Cambridge University Press},\
  \bibinfo {year} {2012})\BibitemShut {NoStop}%
\bibitem [{\citenamefont {Chen}\ \emph {et~al.}(2008)\citenamefont {Chen},
  \citenamefont {Jang}, \citenamefont {Xiao}, \citenamefont {Ishigami},\ and\
  \citenamefont {Fuhrer}}]{Chen2008}%
  \BibitemOpen
  \bibfield  {author} {\bibinfo {author} {\bibfnamefont {J.-H.}\ \bibnamefont
  {Chen}}, \bibinfo {author} {\bibfnamefont {C.}~\bibnamefont {Jang}}, \bibinfo
  {author} {\bibfnamefont {S.}~\bibnamefont {Xiao}}, \bibinfo {author}
  {\bibfnamefont {M.}~\bibnamefont {Ishigami}}, \ and\ \bibinfo {author}
  {\bibfnamefont {M.~S.}\ \bibnamefont {Fuhrer}},\ }\href {\doibase
  10.1038/nnano.2008.58} {\bibfield  {journal} {\bibinfo  {journal} {Nature
  Nanotechnology}\ }\textbf {\bibinfo {volume} {3}},\ \bibinfo {pages} {206}
  (\bibinfo {year} {2008})}\BibitemShut {NoStop}%
\bibitem [{\citenamefont {Efetov}\ and\ \citenamefont
  {Kim}(2010)}]{Efetov2010}%
  \BibitemOpen
  \bibfield  {author} {\bibinfo {author} {\bibfnamefont {D.~K.}\ \bibnamefont
  {Efetov}}\ and\ \bibinfo {author} {\bibfnamefont {P.}~\bibnamefont {Kim}},\
  }\href {http://dx.doi.org/10.1103/PhysRevLett.105.256805} {\bibfield
  {journal} {\bibinfo  {journal} {Phys. Rev. Lett.}\ }\textbf {\bibinfo
  {volume} {105}},\ \bibinfo {pages} {256805} (\bibinfo {year}
  {2010})}\BibitemShut {NoStop}%
\bibitem [{\citenamefont {Graham}\ \emph {et~al.}(2012)\citenamefont {Graham},
  \citenamefont {Shi}, \citenamefont {Ralph}, \citenamefont {Park},\ and\
  \citenamefont {McEuen}}]{Graham2012}%
  \BibitemOpen
  \bibfield  {author} {\bibinfo {author} {\bibfnamefont {M.~W.}\ \bibnamefont
  {Graham}}, \bibinfo {author} {\bibfnamefont {S.-F.}\ \bibnamefont {Shi}},
  \bibinfo {author} {\bibfnamefont {D.~C.}\ \bibnamefont {Ralph}}, \bibinfo
  {author} {\bibfnamefont {J.}~\bibnamefont {Park}}, \ and\ \bibinfo {author}
  {\bibfnamefont {P.~L.}\ \bibnamefont {McEuen}},\ }\href {\doibase
  10.1038/nphys2493} {\bibfield  {journal} {\bibinfo  {journal} {Nature
  Physics}\ }\textbf {\bibinfo {volume} {9}},\ \bibinfo {pages} {103} (\bibinfo
  {year} {2012})}\BibitemShut {NoStop}%
\bibitem [{\citenamefont {Betz}\ \emph
  {et~al.}(2012{\natexlab{b}})\citenamefont {Betz}, \citenamefont {Vialla},
  \citenamefont {Brunel}, \citenamefont {Voisin}, \citenamefont {Picher},
  \citenamefont {Cavanna}, \citenamefont {Madouri}, \citenamefont {F{\`{e}}ve},
  \citenamefont {Berroir}, \citenamefont {Pla{\c{c}}ais},\ and\ \citenamefont
  {Pallecchi}}]{Betz2012a}%
  \BibitemOpen
  \bibfield  {author} {\bibinfo {author} {\bibfnamefont {A.~C.}\ \bibnamefont
  {Betz}}, \bibinfo {author} {\bibfnamefont {F.}~\bibnamefont {Vialla}},
  \bibinfo {author} {\bibfnamefont {D.}~\bibnamefont {Brunel}}, \bibinfo
  {author} {\bibfnamefont {C.}~\bibnamefont {Voisin}}, \bibinfo {author}
  {\bibfnamefont {M.}~\bibnamefont {Picher}}, \bibinfo {author} {\bibfnamefont
  {A.}~\bibnamefont {Cavanna}}, \bibinfo {author} {\bibfnamefont
  {A.}~\bibnamefont {Madouri}}, \bibinfo {author} {\bibfnamefont
  {G.}~\bibnamefont {F{\`{e}}ve}}, \bibinfo {author} {\bibfnamefont {J.-M.}\
  \bibnamefont {Berroir}}, \bibinfo {author} {\bibfnamefont {B.}~\bibnamefont
  {Pla{\c{c}}ais}}, \ and\ \bibinfo {author} {\bibfnamefont {E.}~\bibnamefont
  {Pallecchi}},\ }\href {\doibase 10.1103/PhysRevLett.109.056805} {\bibfield
  {journal} {\bibinfo  {journal} {Physical Review Letters}\ }\textbf {\bibinfo
  {volume} {109}},\ \bibinfo {pages} {056805} (\bibinfo {year}
  {2012}{\natexlab{b}})}\BibitemShut {NoStop}%
\bibitem [{\citenamefont {Graham}\ \emph {et~al.}(2013)\citenamefont {Graham},
  \citenamefont {Shi}, \citenamefont {Wang}, \citenamefont {Ralph},
  \citenamefont {Park},\ and\ \citenamefont {McEuen}}]{Graham2013}%
  \BibitemOpen
  \bibfield  {author} {\bibinfo {author} {\bibfnamefont {M.~W.}\ \bibnamefont
  {Graham}}, \bibinfo {author} {\bibfnamefont {S.-F.}\ \bibnamefont {Shi}},
  \bibinfo {author} {\bibfnamefont {Z.}~\bibnamefont {Wang}}, \bibinfo {author}
  {\bibfnamefont {D.~C.}\ \bibnamefont {Ralph}}, \bibinfo {author}
  {\bibfnamefont {J.}~\bibnamefont {Park}}, \ and\ \bibinfo {author}
  {\bibfnamefont {P.~L.}\ \bibnamefont {McEuen}},\ }\href {\doibase
  10.1021/nl4030787} {\bibfield  {journal} {\bibinfo  {journal} {Nano Letters}\
  }\textbf {\bibinfo {volume} {13}},\ \bibinfo {pages} {5497} (\bibinfo {year}
  {2013})}\BibitemShut {NoStop}%
\bibitem [{\citenamefont {Mariani}\ and\ \citenamefont {von
  Oppen}(2010)}]{Mariani2010}%
  \BibitemOpen
  \bibfield  {author} {\bibinfo {author} {\bibfnamefont {E.}~\bibnamefont
  {Mariani}}\ and\ \bibinfo {author} {\bibfnamefont {F.}~\bibnamefont {von
  Oppen}},\ }\href {\doibase 10.1103/PhysRevB.82.195403} {\bibfield  {journal}
  {\bibinfo  {journal} {Physical Review B}\ }\textbf {\bibinfo {volume} {82}},\
  \bibinfo {pages} {195403} (\bibinfo {year} {2010})}\BibitemShut {NoStop}%
\bibitem [{\citenamefont {Castro}\ \emph {et~al.}(2010)\citenamefont {Castro},
  \citenamefont {Ochoa}, \citenamefont {Katsnelson}, \citenamefont {Gorbachev},
  \citenamefont {Elias}, \citenamefont {Novoselov}, \citenamefont {Geim},\ and\
  \citenamefont {Guinea}}]{Castro2010}%
  \BibitemOpen
  \bibfield  {author} {\bibinfo {author} {\bibfnamefont {E.}~\bibnamefont
  {Castro}}, \bibinfo {author} {\bibfnamefont {H.}~\bibnamefont {Ochoa}},
  \bibinfo {author} {\bibfnamefont {M.}~\bibnamefont {Katsnelson}}, \bibinfo
  {author} {\bibfnamefont {R.}~\bibnamefont {Gorbachev}}, \bibinfo {author}
  {\bibfnamefont {D.}~\bibnamefont {Elias}}, \bibinfo {author} {\bibfnamefont
  {K.}~\bibnamefont {Novoselov}}, \bibinfo {author} {\bibfnamefont
  {A.}~\bibnamefont {Geim}}, \ and\ \bibinfo {author} {\bibfnamefont
  {F.}~\bibnamefont {Guinea}},\ }\href {\doibase
  10.1103/PhysRevLett.105.266601} {\bibfield  {journal} {\bibinfo  {journal}
  {Physical Review Letters}\ }\textbf {\bibinfo {volume} {105}},\ \bibinfo
  {pages} {16} (\bibinfo {year} {2010})}\BibitemShut {NoStop}%
\bibitem [{\citenamefont {Giazotto}\ \emph {et~al.}(2006)\citenamefont
  {Giazotto}, \citenamefont {Heikkil{\"{a}}}, \citenamefont {Luukanen},
  \citenamefont {Savin},\ and\ \citenamefont {Pekola}}]{Giazotto2006}%
  \BibitemOpen
  \bibfield  {author} {\bibinfo {author} {\bibfnamefont {F.}~\bibnamefont
  {Giazotto}}, \bibinfo {author} {\bibfnamefont {T.~T.}\ \bibnamefont
  {Heikkil{\"{a}}}}, \bibinfo {author} {\bibfnamefont {A.}~\bibnamefont
  {Luukanen}}, \bibinfo {author} {\bibfnamefont {A.~M.}\ \bibnamefont {Savin}},
  \ and\ \bibinfo {author} {\bibfnamefont {J.~P.}\ \bibnamefont {Pekola}},\
  }\href {\doibase 10.1103/RevModPhys.78.217} {\bibfield  {journal} {\bibinfo
  {journal} {Rev. Mod. Phys.}\ }\textbf {\bibinfo {volume} {78}},\ \bibinfo
  {pages} {217} (\bibinfo {year} {2006})}\BibitemShut {NoStop}%
\bibitem [{\citenamefont {Fong}\ and\ \citenamefont {Schwab}(2012)}]{Fong2012}%
  \BibitemOpen
  \bibfield  {author} {\bibinfo {author} {\bibfnamefont {K.~C.}\ \bibnamefont
  {Fong}}\ and\ \bibinfo {author} {\bibfnamefont {K.~C.}\ \bibnamefont
  {Schwab}},\ }\href {\doibase 10.1103/PhysRevX.2.031006} {\bibfield  {journal}
  {\bibinfo  {journal} {Physical Review X}\ }\textbf {\bibinfo {volume} {2}},\
  \bibinfo {pages} {031006} (\bibinfo {year} {2012})}\BibitemShut {NoStop}%
\bibitem [{\citenamefont {Fong}\ \emph {et~al.}(2013)\citenamefont {Fong},
  \citenamefont {Wollman}, \citenamefont {Ravi}, \citenamefont {Chen},
  \citenamefont {Clerk}, \citenamefont {Shaw}, \citenamefont {Leduc},\ and\
  \citenamefont {Schwab}}]{Fong2013}%
  \BibitemOpen
  \bibfield  {author} {\bibinfo {author} {\bibfnamefont {K.~C.}\ \bibnamefont
  {Fong}}, \bibinfo {author} {\bibfnamefont {E.~E.}\ \bibnamefont {Wollman}},
  \bibinfo {author} {\bibfnamefont {H.}~\bibnamefont {Ravi}}, \bibinfo {author}
  {\bibfnamefont {W.}~\bibnamefont {Chen}}, \bibinfo {author} {\bibfnamefont
  {A.~A.}\ \bibnamefont {Clerk}}, \bibinfo {author} {\bibfnamefont {M.~D.}\
  \bibnamefont {Shaw}}, \bibinfo {author} {\bibfnamefont {H.~G.}\ \bibnamefont
  {Leduc}}, \ and\ \bibinfo {author} {\bibfnamefont {K.~C.}\ \bibnamefont
  {Schwab}},\ }\href {\doibase 10.1103/PhysRevX.3.041008} {\bibfield  {journal}
  {\bibinfo  {journal} {Physical Review X}\ }\textbf {\bibinfo {volume} {3}},\
  \bibinfo {pages} {041008} (\bibinfo {year} {2013})}\BibitemShut {NoStop}%
\bibitem [{\citenamefont {Han}\ \emph {et~al.}(2013)\citenamefont {Han},
  \citenamefont {Gao}, \citenamefont {Zhang}, \citenamefont {Chen},
  \citenamefont {Chen}, \citenamefont {Liu}, \citenamefont {Zhang},
  \citenamefont {Liu}, \citenamefont {Wu},\ and\ \citenamefont {Yu}}]{Han2013}%
  \BibitemOpen
  \bibfield  {author} {\bibinfo {author} {\bibfnamefont {Q.}~\bibnamefont
  {Han}}, \bibinfo {author} {\bibfnamefont {T.}~\bibnamefont {Gao}}, \bibinfo
  {author} {\bibfnamefont {R.}~\bibnamefont {Zhang}}, \bibinfo {author}
  {\bibfnamefont {Y.}~\bibnamefont {Chen}}, \bibinfo {author} {\bibfnamefont
  {J.}~\bibnamefont {Chen}}, \bibinfo {author} {\bibfnamefont {G.}~\bibnamefont
  {Liu}}, \bibinfo {author} {\bibfnamefont {Y.}~\bibnamefont {Zhang}}, \bibinfo
  {author} {\bibfnamefont {Z.}~\bibnamefont {Liu}}, \bibinfo {author}
  {\bibfnamefont {X.}~\bibnamefont {Wu}}, \ and\ \bibinfo {author}
  {\bibfnamefont {D.}~\bibnamefont {Yu}},\ }\href {\doibase 10.1038/srep03533}
  {\bibfield  {journal} {\bibinfo  {journal} {Scientific Reports}\ }\textbf
  {\bibinfo {volume} {3}},\ \bibinfo {pages} {3533} (\bibinfo {year}
  {2013})}\BibitemShut {NoStop}%
\bibitem [{\citenamefont {Vora}\ \emph {et~al.}(2014)\citenamefont {Vora},
  \citenamefont {Nielsen},\ and\ \citenamefont {Du}}]{Vora2014}%
  \BibitemOpen
  \bibfield  {author} {\bibinfo {author} {\bibfnamefont {H.}~\bibnamefont
  {Vora}}, \bibinfo {author} {\bibfnamefont {B.}~\bibnamefont {Nielsen}}, \
  and\ \bibinfo {author} {\bibfnamefont {X.}~\bibnamefont {Du}},\ }\href
  {\doibase 10.1063/1.4866325} {\bibfield  {journal} {\bibinfo  {journal}
  {Journal of Applied Physics}\ }\textbf {\bibinfo {volume} {115}},\ \bibinfo
  {pages} {074505} (\bibinfo {year} {2014})}\BibitemShut {NoStop}%
\bibitem [{\citenamefont {McKitterick}\ \emph {et~al.}(2016)\citenamefont
  {McKitterick}, \citenamefont {Prober},\ and\ \citenamefont
  {Rooks}}]{McKitterick2016}%
  \BibitemOpen
  \bibfield  {author} {\bibinfo {author} {\bibfnamefont {C.~B.}\ \bibnamefont
  {McKitterick}}, \bibinfo {author} {\bibfnamefont {D.~E.}\ \bibnamefont
  {Prober}}, \ and\ \bibinfo {author} {\bibfnamefont {M.~J.}\ \bibnamefont
  {Rooks}},\ }\href {\doibase 10.1103/PhysRevB.93.075410} {\bibfield  {journal}
  {\bibinfo  {journal} {Physical Review B}\ }\textbf {\bibinfo {volume} {93}},\
  \bibinfo {pages} {075410} (\bibinfo {year} {2016})},\ \Eprint
  {http://arxiv.org/abs/1505.07034} {arXiv:1505.07034} \BibitemShut {NoStop}%
\bibitem [{\citenamefont {{El Fatimy}}\ \emph {et~al.}(2019)\citenamefont {{El
  Fatimy}}, \citenamefont {Han}, \citenamefont {Quirk}, \citenamefont {{St.
  Marie}}, \citenamefont {Dejarld}, \citenamefont {Myers-Ward}, \citenamefont
  {Daniels}, \citenamefont {Pavunny}, \citenamefont {Gaskill}, \citenamefont
  {Aytac}, \citenamefont {Murphy},\ and\ \citenamefont
  {Barbara}}]{ElFatimy2019}%
  \BibitemOpen
  \bibfield  {author} {\bibinfo {author} {\bibfnamefont {A.}~\bibnamefont {{El
  Fatimy}}}, \bibinfo {author} {\bibfnamefont {P.}~\bibnamefont {Han}},
  \bibinfo {author} {\bibfnamefont {N.}~\bibnamefont {Quirk}}, \bibinfo
  {author} {\bibfnamefont {L.}~\bibnamefont {{St. Marie}}}, \bibinfo {author}
  {\bibfnamefont {M.~T.}\ \bibnamefont {Dejarld}}, \bibinfo {author}
  {\bibfnamefont {R.~L.}\ \bibnamefont {Myers-Ward}}, \bibinfo {author}
  {\bibfnamefont {K.}~\bibnamefont {Daniels}}, \bibinfo {author} {\bibfnamefont
  {S.}~\bibnamefont {Pavunny}}, \bibinfo {author} {\bibfnamefont {D.~K.}\
  \bibnamefont {Gaskill}}, \bibinfo {author} {\bibfnamefont {Y.}~\bibnamefont
  {Aytac}}, \bibinfo {author} {\bibfnamefont {T.~E.}\ \bibnamefont {Murphy}}, \
  and\ \bibinfo {author} {\bibfnamefont {P.}~\bibnamefont {Barbara}},\ }\href
  {\doibase 10.1016/J.CARBON.2019.08.019} {\bibfield  {journal} {\bibinfo
  {journal} {Carbon}\ }\textbf {\bibinfo {volume} {154}},\ \bibinfo {pages}
  {497} (\bibinfo {year} {2019})}\BibitemShut {NoStop}%
\bibitem [{\citenamefont {Courtois}\ \emph {et~al.}(2008)\citenamefont
  {Courtois}, \citenamefont {Meschke}, \citenamefont {Peltonen},\ and\
  \citenamefont {Pekola}}]{Courtois2008}%
  \BibitemOpen
  \bibfield  {author} {\bibinfo {author} {\bibfnamefont {H.}~\bibnamefont
  {Courtois}}, \bibinfo {author} {\bibfnamefont {M.}~\bibnamefont {Meschke}},
  \bibinfo {author} {\bibfnamefont {J.~T.}\ \bibnamefont {Peltonen}}, \ and\
  \bibinfo {author} {\bibfnamefont {J.~P.}\ \bibnamefont {Pekola}},\ }\href
  {\doibase 10.1103/PhysRevLett.101.067002} {\bibfield  {journal} {\bibinfo
  {journal} {Physical Review Letters}\ }\textbf {\bibinfo {volume} {101}},\
  \bibinfo {pages} {067002} (\bibinfo {year} {2008})}\BibitemShut {NoStop}%
\bibitem [{\citenamefont {Segev}\ \emph {et~al.}(2007)\citenamefont {Segev},
  \citenamefont {Abdo}, \citenamefont {Shtempluck},\ and\ \citenamefont
  {Buks}}]{Segev_2007}%
  \BibitemOpen
  \bibfield  {author} {\bibinfo {author} {\bibfnamefont {E.}~\bibnamefont
  {Segev}}, \bibinfo {author} {\bibfnamefont {B.}~\bibnamefont {Abdo}},
  \bibinfo {author} {\bibfnamefont {O.}~\bibnamefont {Shtempluck}}, \ and\
  \bibinfo {author} {\bibfnamefont {E.}~\bibnamefont {Buks}},\ }\href {\doibase
  10.1088/0953-8984/19/9/096206} {\bibfield  {journal} {\bibinfo  {journal}
  {Journal of Physics: Condensed Matter}\ }\textbf {\bibinfo {volume} {19}},\
  \bibinfo {pages} {096206} (\bibinfo {year} {2007})}\BibitemShut {NoStop}%
\bibitem [{\citenamefont {Madiot}\ \emph {et~al.}(2021)\citenamefont {Madiot},
  \citenamefont {Barbay},\ and\ \citenamefont
  {Braive}}]{madiot_vibrational_2021}%
  \BibitemOpen
  \bibfield  {author} {\bibinfo {author} {\bibfnamefont {G.}~\bibnamefont
  {Madiot}}, \bibinfo {author} {\bibfnamefont {S.}~\bibnamefont {Barbay}}, \
  and\ \bibinfo {author} {\bibfnamefont {R.}~\bibnamefont {Braive}},\ }\href
  {\doibase 10.1021/acs.nanolett.1c02879} {\bibfield  {journal} {\bibinfo
  {journal} {Nano Letters}\ }\textbf {\bibinfo {volume} {21}},\ \bibinfo
  {pages} {8311} (\bibinfo {year} {2021})}\BibitemShut {NoStop}%
\bibitem [{\citenamefont {Urgell}\ \emph {et~al.}(2020)\citenamefont {Urgell},
  \citenamefont {Yang}, \citenamefont {De~Bonis}, \citenamefont {Samanta},
  \citenamefont {Esplandiu}, \citenamefont {Dong}, \citenamefont {Jin},\ and\
  \citenamefont {Bachtold}}]{urgell_cooling_2020}%
  \BibitemOpen
  \bibfield  {author} {\bibinfo {author} {\bibfnamefont {C.}~\bibnamefont
  {Urgell}}, \bibinfo {author} {\bibfnamefont {W.}~\bibnamefont {Yang}},
  \bibinfo {author} {\bibfnamefont {S.~L.}\ \bibnamefont {De~Bonis}}, \bibinfo
  {author} {\bibfnamefont {C.}~\bibnamefont {Samanta}}, \bibinfo {author}
  {\bibfnamefont {M.~J.}\ \bibnamefont {Esplandiu}}, \bibinfo {author}
  {\bibfnamefont {Q.}~\bibnamefont {Dong}}, \bibinfo {author} {\bibfnamefont
  {Y.}~\bibnamefont {Jin}}, \ and\ \bibinfo {author} {\bibfnamefont
  {A.}~\bibnamefont {Bachtold}},\ }\href {\doibase 10.1038/s41567-019-0682-6}
  {\bibfield  {journal} {\bibinfo  {journal} {Nature Physics}\ }\textbf
  {\bibinfo {volume} {16}},\ \bibinfo {pages} {32} (\bibinfo {year}
  {2020})}\BibitemShut {NoStop}%
\bibitem [{\citenamefont {Suret}\ \emph {et~al.}(2000)\citenamefont {Suret},
  \citenamefont {Derozier}, \citenamefont {Lefranc}, \citenamefont {Zemmouri},\
  and\ \citenamefont {Bielawski}}]{Suret_2000}%
  \BibitemOpen
  \bibfield  {author} {\bibinfo {author} {\bibfnamefont {P.}~\bibnamefont
  {Suret}}, \bibinfo {author} {\bibfnamefont {D.}~\bibnamefont {Derozier}},
  \bibinfo {author} {\bibfnamefont {M.}~\bibnamefont {Lefranc}}, \bibinfo
  {author} {\bibfnamefont {J.}~\bibnamefont {Zemmouri}}, \ and\ \bibinfo
  {author} {\bibfnamefont {S.}~\bibnamefont {Bielawski}},\ }\href {\doibase
  10.1103/PhysRevA.61.021805} {\bibfield  {journal} {\bibinfo  {journal} {Phys.
  Rev. A}\ }\textbf {\bibinfo {volume} {61}},\ \bibinfo {pages} {021805}
  (\bibinfo {year} {2000})}\BibitemShut {NoStop}%
\bibitem [{\citenamefont {Hao}\ \emph {et~al.}(2014)\citenamefont {Hao},
  \citenamefont {Rouxinol},\ and\ \citenamefont {LaHaye}}]{HaoYu2014}%
  \BibitemOpen
  \bibfield  {author} {\bibinfo {author} {\bibfnamefont {Y.}~\bibnamefont
  {Hao}}, \bibinfo {author} {\bibfnamefont {F.}~\bibnamefont {Rouxinol}}, \
  and\ \bibinfo {author} {\bibfnamefont {M.~D.}\ \bibnamefont {LaHaye}},\
  }\href {\doibase 10.1063/1.4903777} {\bibfield  {journal} {\bibinfo
  {journal} {Applied Physics Letters}\ }\textbf {\bibinfo {volume} {105}},\
  \bibinfo {pages} {222603} (\bibinfo {year} {2014})},\ \Eprint
  {http://arxiv.org/abs/https://doi.org/10.1063/1.4903777}
  {https://doi.org/10.1063/1.4903777} \BibitemShut {NoStop}%
\bibitem [{\citenamefont {Kumar}\ \emph {et~al.}(2018)\citenamefont {Kumar},
  \citenamefont {Laitinen},\ and\ \citenamefont {Hakonen}}]{Kumar2018}%
  \BibitemOpen
  \bibfield  {author} {\bibinfo {author} {\bibfnamefont {M.}~\bibnamefont
  {Kumar}}, \bibinfo {author} {\bibfnamefont {A.}~\bibnamefont {Laitinen}}, \
  and\ \bibinfo {author} {\bibfnamefont {P.}~\bibnamefont {Hakonen}},\ }\href
  {\doibase 10.1038/s41467-018-05094-8} {\bibfield  {journal} {\bibinfo
  {journal} {Nature Communications}\ }\textbf {\bibinfo {volume} {9}},\
  \bibinfo {pages} {2776} (\bibinfo {year} {2018})},\ \Eprint
  {http://arxiv.org/abs/1611.02742} {arXiv:1611.02742} \BibitemShut {NoStop}%
\bibitem [{\citenamefont {Kamada}\ \emph {et~al.}(2021)\citenamefont {Kamada},
  \citenamefont {Gall}, \citenamefont {Sarkar}, \citenamefont {Kumar},
  \citenamefont {Laitinen}, \citenamefont {Gornyi},\ and\ \citenamefont
  {Hakonen}}]{Kamada2021}%
  \BibitemOpen
  \bibfield  {author} {\bibinfo {author} {\bibfnamefont {M.}~\bibnamefont
  {Kamada}}, \bibinfo {author} {\bibfnamefont {V.}~\bibnamefont {Gall}},
  \bibinfo {author} {\bibfnamefont {J.}~\bibnamefont {Sarkar}}, \bibinfo
  {author} {\bibfnamefont {M.}~\bibnamefont {Kumar}}, \bibinfo {author}
  {\bibfnamefont {A.}~\bibnamefont {Laitinen}}, \bibinfo {author}
  {\bibfnamefont {I.}~\bibnamefont {Gornyi}}, \ and\ \bibinfo {author}
  {\bibfnamefont {P.}~\bibnamefont {Hakonen}},\ }\href {\doibase
  10.1103/PhysRevB.104.115432} {\bibfield  {journal} {\bibinfo  {journal}
  {Phys. Rev. B}\ }\textbf {\bibinfo {volume} {104}},\ \bibinfo {pages}
  {115432} (\bibinfo {year} {2021})}\BibitemShut {NoStop}%
\bibitem [{\citenamefont {Couto}\ \emph {et~al.}(2014)\citenamefont {Couto},
  \citenamefont {Costanzo}, \citenamefont {Engels}, \citenamefont {Ki},
  \citenamefont {Watanabe}, \citenamefont {Taniguchi}, \citenamefont
  {Stampfer}, \citenamefont {Guinea},\ and\ \citenamefont
  {Morpurgo}}]{Cuoto2014}%
  \BibitemOpen
  \bibfield  {author} {\bibinfo {author} {\bibfnamefont {N.~J.~G.}\
  \bibnamefont {Couto}}, \bibinfo {author} {\bibfnamefont {D.}~\bibnamefont
  {Costanzo}}, \bibinfo {author} {\bibfnamefont {S.}~\bibnamefont {Engels}},
  \bibinfo {author} {\bibfnamefont {D.-K.}\ \bibnamefont {Ki}}, \bibinfo
  {author} {\bibfnamefont {K.}~\bibnamefont {Watanabe}}, \bibinfo {author}
  {\bibfnamefont {T.}~\bibnamefont {Taniguchi}}, \bibinfo {author}
  {\bibfnamefont {C.}~\bibnamefont {Stampfer}}, \bibinfo {author}
  {\bibfnamefont {F.}~\bibnamefont {Guinea}}, \ and\ \bibinfo {author}
  {\bibfnamefont {A.~F.}\ \bibnamefont {Morpurgo}},\ }\href {\doibase
  10.1103/PhysRevX.4.041019} {\bibfield  {journal} {\bibinfo  {journal} {Phys.
  Rev. X}\ }\textbf {\bibinfo {volume} {4}},\ \bibinfo {pages} {041019}
  (\bibinfo {year} {2014})}\BibitemShut {NoStop}%
\bibitem [{\citenamefont {Borzenets}\ \emph {et~al.}(2013)\citenamefont
  {Borzenets}, \citenamefont {Coskun}, \citenamefont {Mebrahtu}, \citenamefont
  {Bomze}, \citenamefont {Smirnov},\ and\ \citenamefont
  {Finkelstein}}]{Borzenets2013}%
  \BibitemOpen
  \bibfield  {author} {\bibinfo {author} {\bibfnamefont {I.~V.}\ \bibnamefont
  {Borzenets}}, \bibinfo {author} {\bibfnamefont {U.~C.}\ \bibnamefont
  {Coskun}}, \bibinfo {author} {\bibfnamefont {H.~T.}\ \bibnamefont
  {Mebrahtu}}, \bibinfo {author} {\bibfnamefont {Y.~V.}\ \bibnamefont {Bomze}},
  \bibinfo {author} {\bibfnamefont {A.~I.}\ \bibnamefont {Smirnov}}, \ and\
  \bibinfo {author} {\bibfnamefont {G.}~\bibnamefont {Finkelstein}},\ }\href
  {\doibase 10.1103/PhysRevLett.111.027001} {\bibfield  {journal} {\bibinfo
  {journal} {Physical Review Letters}\ }\textbf {\bibinfo {volume} {111}},\
  \bibinfo {pages} {027001} (\bibinfo {year} {2013})}\BibitemShut {NoStop}%
\bibitem [{\citenamefont {Khomyakov}\ \emph {et~al.}(2009)\citenamefont
  {Khomyakov}, \citenamefont {Giovannetti}, \citenamefont {Rusu}, \citenamefont
  {Brocks}, \citenamefont {van~den Brink},\ and\ \citenamefont
  {Kelly}}]{Khomyakov2009}%
  \BibitemOpen
  \bibfield  {author} {\bibinfo {author} {\bibfnamefont {P.~A.}\ \bibnamefont
  {Khomyakov}}, \bibinfo {author} {\bibfnamefont {G.}~\bibnamefont
  {Giovannetti}}, \bibinfo {author} {\bibfnamefont {P.~C.}\ \bibnamefont
  {Rusu}}, \bibinfo {author} {\bibfnamefont {G.}~\bibnamefont {Brocks}},
  \bibinfo {author} {\bibfnamefont {J.}~\bibnamefont {van~den Brink}}, \ and\
  \bibinfo {author} {\bibfnamefont {P.~J.}\ \bibnamefont {Kelly}},\ }\href
  {\doibase 10.1103/PhysRevB.79.195425} {\bibfield  {journal} {\bibinfo
  {journal} {Physical Review B}\ }\textbf {\bibinfo {volume} {79}},\ \bibinfo
  {pages} {195425} (\bibinfo {year} {2009})}\BibitemShut {NoStop}%
\bibitem [{\citenamefont {Laitinen}\ \emph {et~al.}(2016)\citenamefont
  {Laitinen}, \citenamefont {Paraoanu}, \citenamefont {Oksanen}, \citenamefont
  {Craciun}, \citenamefont {Russo}, \citenamefont {Sonin},\ and\ \citenamefont
  {Hakonen}}]{Laitinen2016}%
  \BibitemOpen
  \bibfield  {author} {\bibinfo {author} {\bibfnamefont {A.}~\bibnamefont
  {Laitinen}}, \bibinfo {author} {\bibfnamefont {G.~S.}\ \bibnamefont
  {Paraoanu}}, \bibinfo {author} {\bibfnamefont {M.}~\bibnamefont {Oksanen}},
  \bibinfo {author} {\bibfnamefont {M.~F.}\ \bibnamefont {Craciun}}, \bibinfo
  {author} {\bibfnamefont {S.}~\bibnamefont {Russo}}, \bibinfo {author}
  {\bibfnamefont {E.}~\bibnamefont {Sonin}}, \ and\ \bibinfo {author}
  {\bibfnamefont {P.}~\bibnamefont {Hakonen}},\ }\href {\doibase
  10.1103/PhysRevB.93.115413} {\bibfield  {journal} {\bibinfo  {journal}
  {Physical Review B}\ }\textbf {\bibinfo {volume} {93}},\ \bibinfo {pages} {1}
  (\bibinfo {year} {2016})},\ \Eprint {http://arxiv.org/abs/1502.04330v2}
  {arXiv:1502.04330v2} \BibitemShut {NoStop}%
\bibitem [{\citenamefont {Nagashio}\ \emph {et~al.}(2009)\citenamefont
  {Nagashio}, \citenamefont {Nishimura}, \citenamefont {Kita},\ and\
  \citenamefont {Toriumi}}]{Nagashio2009}%
  \BibitemOpen
  \bibfield  {author} {\bibinfo {author} {\bibfnamefont {K.}~\bibnamefont
  {Nagashio}}, \bibinfo {author} {\bibfnamefont {T.}~\bibnamefont {Nishimura}},
  \bibinfo {author} {\bibfnamefont {K.}~\bibnamefont {Kita}}, \ and\ \bibinfo
  {author} {\bibfnamefont {A.}~\bibnamefont {Toriumi}},\ }\href {\doibase
  10.1109/IEDM.2009.5424297} {\bibfield  {journal} {\bibinfo  {journal} {2009
  IEEE International Electron Devices Meeting (IEDM)}\ }\textbf {\bibinfo
  {volume} {23}},\ \bibinfo {pages} {2.1} (\bibinfo {year} {2009})}\BibitemShut
  {NoStop}%
\bibitem [{\citenamefont {Aamir}\ \emph {et~al.}(2021)\citenamefont {Aamir},
  \citenamefont {Moore}, \citenamefont {Lu}, \citenamefont {Seifert},
  \citenamefont {Englund}, \citenamefont {Fong},\ and\ \citenamefont
  {Efetov}}]{Aamir2021}%
  \BibitemOpen
  \bibfield  {author} {\bibinfo {author} {\bibfnamefont {M.~A.}\ \bibnamefont
  {Aamir}}, \bibinfo {author} {\bibfnamefont {J.~N.}\ \bibnamefont {Moore}},
  \bibinfo {author} {\bibfnamefont {X.}~\bibnamefont {Lu}}, \bibinfo {author}
  {\bibfnamefont {P.}~\bibnamefont {Seifert}}, \bibinfo {author} {\bibfnamefont
  {D.}~\bibnamefont {Englund}}, \bibinfo {author} {\bibfnamefont {K.~C.}\
  \bibnamefont {Fong}}, \ and\ \bibinfo {author} {\bibfnamefont {D.~K.}\
  \bibnamefont {Efetov}},\ }\href {\doibase 10.1021/acs.nanolett.1c01553}
  {\bibfield  {journal} {\bibinfo  {journal} {Nano Letters}\ }\textbf {\bibinfo
  {volume} {21}},\ \bibinfo {pages} {5330} (\bibinfo {year}
  {2021})}\BibitemShut {NoStop}%
\bibitem [{\citenamefont {Tomi}\ \emph {et~al.}(2021)\citenamefont {Tomi},
  \citenamefont {Samatov}, \citenamefont {Vasenko}, \citenamefont {Laitinen},
  \citenamefont {Hakonen},\ and\ \citenamefont {Golubev}}]{Tomi2021}%
  \BibitemOpen
  \bibfield  {author} {\bibinfo {author} {\bibfnamefont {M.}~\bibnamefont
  {Tomi}}, \bibinfo {author} {\bibfnamefont {M.~R.}\ \bibnamefont {Samatov}},
  \bibinfo {author} {\bibfnamefont {A.~S.}\ \bibnamefont {Vasenko}}, \bibinfo
  {author} {\bibfnamefont {A.}~\bibnamefont {Laitinen}}, \bibinfo {author}
  {\bibfnamefont {P.}~\bibnamefont {Hakonen}}, \ and\ \bibinfo {author}
  {\bibfnamefont {D.~S.}\ \bibnamefont {Golubev}},\ }\href {\doibase
  10.1103/PhysRevB.104.134513} {\bibfield  {journal} {\bibinfo  {journal}
  {Physical Review B}\ }\textbf {\bibinfo {volume} {104}},\ \bibinfo {pages}
  {134513} (\bibinfo {year} {2021})}\BibitemShut {NoStop}%
\bibitem [{\citenamefont {Voutilainen}\ \emph {et~al.}(2011)\citenamefont
  {Voutilainen}, \citenamefont {Fay}, \citenamefont {H\"akkinen}, \citenamefont
  {Viljas}, \citenamefont {Heikkil\"a},\ and\ \citenamefont
  {Hakonen}}]{Voutilainen2011}%
  \BibitemOpen
  \bibfield  {author} {\bibinfo {author} {\bibfnamefont {J.}~\bibnamefont
  {Voutilainen}}, \bibinfo {author} {\bibfnamefont {A.}~\bibnamefont {Fay}},
  \bibinfo {author} {\bibfnamefont {P.}~\bibnamefont {H\"akkinen}}, \bibinfo
  {author} {\bibfnamefont {J.~K.}\ \bibnamefont {Viljas}}, \bibinfo {author}
  {\bibfnamefont {T.~T.}\ \bibnamefont {Heikkil\"a}}, \ and\ \bibinfo {author}
  {\bibfnamefont {P.~J.}\ \bibnamefont {Hakonen}},\ }\href {\doibase
  10.1103/PhysRevB.84.045419} {\bibfield  {journal} {\bibinfo  {journal}
  {Physical Review B}\ }\textbf {\bibinfo {volume} {84}},\ \bibinfo {pages}
  {045419} (\bibinfo {year} {2011})}\BibitemShut {NoStop}%
\end{thebibliography}%

\end{document}